\documentclass{aa}
\usepackage{graphicx}
\usepackage{txfonts}
\usepackage{booktabs}
\usepackage{wrapfig}
\usepackage[version=4]{mhchem}
\usepackage{color} 
\newcommand{\SgrB}{Sgr\,B2}

\newcommand{\thiform}{\ce{H2CS} }

\newcommand{\phn}{\phantom{0}}
\newcommand{\phnn}{\phantom{0}\phantom{0}}
\newcommand{\phnx}{\phantom{\tablefootmark{b}}}

\begin{document}
	
\title{The Physical and chemical structure of Sagittarius~B2}
\subtitle{IV. Converging filaments in the high-mass cluster forming region
Sgr\,B2(N)}
\author{A.~Schw\"orer\inst{1}
\and \'A.~S\'anchez-Monge\inst{1}
\and P.~Schilke\inst{1}
\and T.~Möller\inst{1}
\and A.~Ginsburg\inst{2}
\and F.~Meng\inst{1}
\and A.~Schmiedeke\inst{3}
\and H.~S.~P.~M\"uller\inst{1}
\and D.~Lis\inst{4}
\and S.~-L.~Qin\inst{5}
}
\authorrunning{A. Schw\"orer, \'A. S\'anchez-Monge, P. Schilke et al.}
\institute{I. Physikalisches Institut, Universit\"at zu K\"oln,
Z\"ulpicher Strasse 77, 50937 K\"oln, Germany \\ \email{schwoerer@ph1.uni-koeln.de}
\and National Radio Astronomy Observatory, 1003 Lopezville Rd., Socorro,
NM 87801, USA
\and Max-Planck-Institut für extraterrestrische Physik, Gießenbachstraße
1, 85748 Garching bei M\"unchen, Germany
\and Jet Propulsion Laboratory, California Institute of Technology, 4800 Oak Grove Drive, Pasadena, CA 91109, USA \\
Sorbonne Universit\'{e}, Observatoire de Paris, Universit\'{e} PSL,
CNRS, LERMA, F-75005, Paris, France
\and Department of Astronomy, Yunnan University, and Key Laboratory
of Astroparticle Physics of Yunnan Province, Kumming,
650091, China}
\date{Received, accepted}

\abstract
{Sagittarius B2 (north) is a chemically rich, high-mass star-forming region
located within the giant molecular cloud complex \SgrB\ in the central molecular
zone of our Galaxy. Dust continuum emission at 242~GHz, obtained at high
angular resolution with the Atacama Large Millimeter Array (ALMA), reveals
that it has a filamentary structure on scales of 0.1~pc.} %, with high densities of 10$^8$--10$^9$~cm$^{-3}$.
{We aim to characterize the filamentary structure of \SgrB(N) and its kinematic
properties using multiple molecular dense gas tracers.}
{We have used an unbiased, spectral line-survey that covers the frequency
range from 211 to 275~GHz and obtained with ALMA (angular resolution of 0$\farcs$4,
or 3300~au) to study the small-scale structure of the dense gas in \SgrB(N).
In order to derive the kinematic properties of the gas in a chemically line-rich
source like \SgrB(N), we have developed a python-based tool that stacks all
the detected line transitions of any molecular species. This allows us to
increase the signal-to-noise ratio of our observations and average out line
blending effects, which are common in line-rich regions.}
{A filamentary network is visible in \SgrB(N) in the emission maps of the
molecular species \ce{CH3OCHO}, \ce{CH3OCH3}, \ce{CH3OH} and \ce{H2CS}. In
total, eight filaments are found that converge to the central hub (with a
mass of 2000~$M_\odot$, assuming a temperature of 250~K) and extending for
about 0.1~pc (up to 0.5~pc). The spatial structure, together with the presence
of the massive central region, suggest that these filaments may be associated
with accretion processes, transporting material from the outer regions to
the central dense hub. We derive velocity gradients along the filaments of
about 20--100~km~s$^{-1}$~pc$^{-1}$, which are 10--100 times larger than
those typically found on larger scales ($\sim$1~pc) in other star-forming
regions. The mass accretion rates of individual filaments are $\lessapprox$
0.05~M$_\odot$~yr$^{-1}$, which result in a total accretion rate of 0.16~$M_\odot$~yr$^{-1}$.
Some filaments harbor dense cores that are likely forming stars and stellar
clusters. We determine an empirical relation between the luminosity and stellar
mass of the clusters. The stellar content of these dense cores is on the
order of 50\% of the total mass. The timescales required for the dense cores
to collapse and form stars, exhausting their gas content, are compared with
the timescale of their accretion process onto the central hub. We conclude
that the cores may merge in the center when already forming stellar clusters
but still containing a significant amount of gas, resulting in a "damp"
merger.}
{The high density and mass of the central region, combined with the presence
of converging filaments with high mass, high accretion rates and embedded
dense cores already forming stars, suggest that \SgrB(N) may have the potential
to evolve into a super stellar cluster.}

\keywords{stars: formation --- stars: massive --- techniques: high angular
resolution --- radio lines: ISM --- ISM: kinematics and dynamics --- ISM:
individual objects \SgrB(N)}

	\maketitle
%
%-------------------------------------------------------------------
%--------------------------------------------------------------------
\section{Introduction}\label{sec:intro}
The star-forming complex Sagittarius B2 (hereafter \SgrB) is the most massive
cloud in our Galaxy with a mass of $~10^7$~M$_\odot$ and H$_2$ densities
of 10$^3$--10$^5$~cm$^{-3}$ (\citealt{Schmiedeke2016}, \citealt{Huettemeister1995},
\citealt{Goldsmith1990}), located at a distance\footnote{For consistency
with previous findings in the literature, for our calculations we used the
distance $8.34 \pm 0.16$~kpc \citep{Reid2014}} of $8.127 \pm 0.031$~kpc
\citep{GravityCollaboration2018}
in the vicinity of the Galactic center (at a projected distance of 107~pc)
within the Central Molecular Zone (CMZ). The CMZ provides an extreme environment
in terms of pressure, turbulent Mach number, and gas temperature ($\sim$60
to >100~K), which are much higher than those found in star-forming regions
distributed throughout the Galactic disk (Ginsburg et. al 2016, \citealt{Morris1996}),
but comparable to the physical conditions found in starburst galaxies. Therefore,
the CMZ and \SgrB\, are perfect targets to study star formation under extreme
conditions in our local environment.
The complex
\SgrB~harbors two main sites of active high-mass star formation, \SgrB~Main~(M)
and North~(N), with masses of 10$^4$--10$^5$~M$_\odot$ and \ce{H2} densities
of $\sim$10$^7$~cm$^{-3}$. Moreover, active high-mass star formation and
a huge number of different (complex) molecules have been observed in both
of them \citep[e.g.][]
{DePree2014,Belloche2013,Schmiedeke2016,Sanchez2017,Ginsburg2018,Mills2018}.
We have begun a project characterizing the physical and chemical properties
of \SgrB\, covering all the spatial scales from au up to tens of parsec.
For this, we combined multiwavelength observations with 3D radiative transfer
modeling (see \citealt{Schmiedeke2016,Sanchez2017,Pols2018}). In this paper,
the fourth of the series, we make use of high-spatial resolution observations
obtained with ALMA\footnote{Atacama Large Millimeter/submillimeter Array,
ALMA partnership et al. 2015} (0$\farcs$4, or 3300 au) to study the small-scale
structure of the dense gas in \SgrB(N). The dust continuum emission studied
in \citet[][herafter Paper II]{Sanchez2017} revealed a filamentary network
converging to a central hub, which has a total mass of about 2000~$M_\odot$
(assuming a temperature of 250 K)
and densities $\sim$10$^9$ cm$^{-3}$. In this paper, we characterized the
kinematic properties of the filaments in
more detail by making use of the molecular line emission observed with ALMA.
Filaments have been found, in part thanks to the \textit{Herschel} satellite,
to be ubiquitous entities in molecular clouds at many different scales. Their
presence in almost all the environments suggest that they play an important
role during the early phases of the star formation process. It has been found
in local low-mass star forming regions at scales of a few pc that filaments
are hatcheries of prestellar cores (e.g. Taurus B211/B213 , \citealt{Palmeirim2013,Marsh2016}).
Similarly, molecular clouds forming high-mass stars (O and B spectral type
stars) are found to be pervaded by large and complex filamentary networks,
when observed at large scales of a few pc (see e.g.\ Orion: \citealt{Suri2019,Hacar2018};
MonR2: Trevino-Morales et al. in prep). In these cases, stellar clusters
are found to form in dense hubs with densities $>$10$^6$~cm$^{-3}$. Denser
high-mass star forming regions like W33A ($\sim$~10$^7$-10$^8$~cm$^{-3}$),
when observed with high enough angular resolution, show filamentary structures
at scales of a few 1000~au and converging toward the most massive objects
forming in the clusters (\citealt{Maud2017,Izquierdo2018}). \SgrB(N) provides
us, thanks to the high densities of 10$^8$-10$^9$~cm$^{-3}$, with an ideal
target to characterize filaments in even more extreme environments. Moreover,
the high densities and masses found in the central region of \SgrB(N) suggest
that this region may evolve into a super stellar cluster or young massive
cluster (YMC), similar to the Arches and Quintuplet clusters. YMCs are considered
to be predecessors of globular clusters, which are formed at the earliest
epochs of the universe. Their formation process is still poorly understood,
in particular, the mechanism of mass accumulation, during which filaments
may play an important role.
This paper is structured as follows. In Sect. 2 we describe our
ALMA observations. In Sect. 3 we present the physical and kinematic properties
of the filaments found in \SgrB(N). In Sect. 4 we discuss our results of
the converging filaments and argue that \SgrB(N) may form a super stellar
cluster. Finally, in Sect. 5 we summarize the main results.
%--------------------------------------------------------------------
\section{ALMA observations}\label{sec:obs}
In June 2014 and June 2015, \SgrB~was observed with ALMA (Atacama Large
Millimeter/submillimeter Array) during its Cycle 2. 34--36 antennas were
used in an extended configuration with baselines in the range from 30 to
650~m, which provided an angular resolution of 0$\farcs$3--0$\farcs$7.
The chosen spectral scan mode surveyed the whole ALMA band~6 (211 to 275~GHz)
with 10 different spectral tunings, providing a spectral resolution of 0.5--0.7~km~s$^{-1}$.
\SgrB(N), with phase center at $\alpha$(J2000)=17$^{h}$47$^{m}$19$^{s}$.887,
$\delta$(J2000)=$-$28$^\circ$22$'$15\farcs 76, was observed in track-sharing
mode together with \SgrB(M). The calibration and imaging processes were carried
out with CASA\footnote{The Common Astronomy Software Applications (CASA;
McMullin
et al. 2007). Downloaded at http://casa.nrao.edu}
version 4.4.0. All the images were restored with a common Gaussian beam of
0\farcs4. Further details of the observations, calibration, and imaging are
provided in Paper~II.
%--------------------------------------------------------------------\methform
\section{Filamentary structure in \SgrB(N)}\label{sec:results}
\begin{figure}
\centering
\includegraphics[width=0.95\columnwidth]{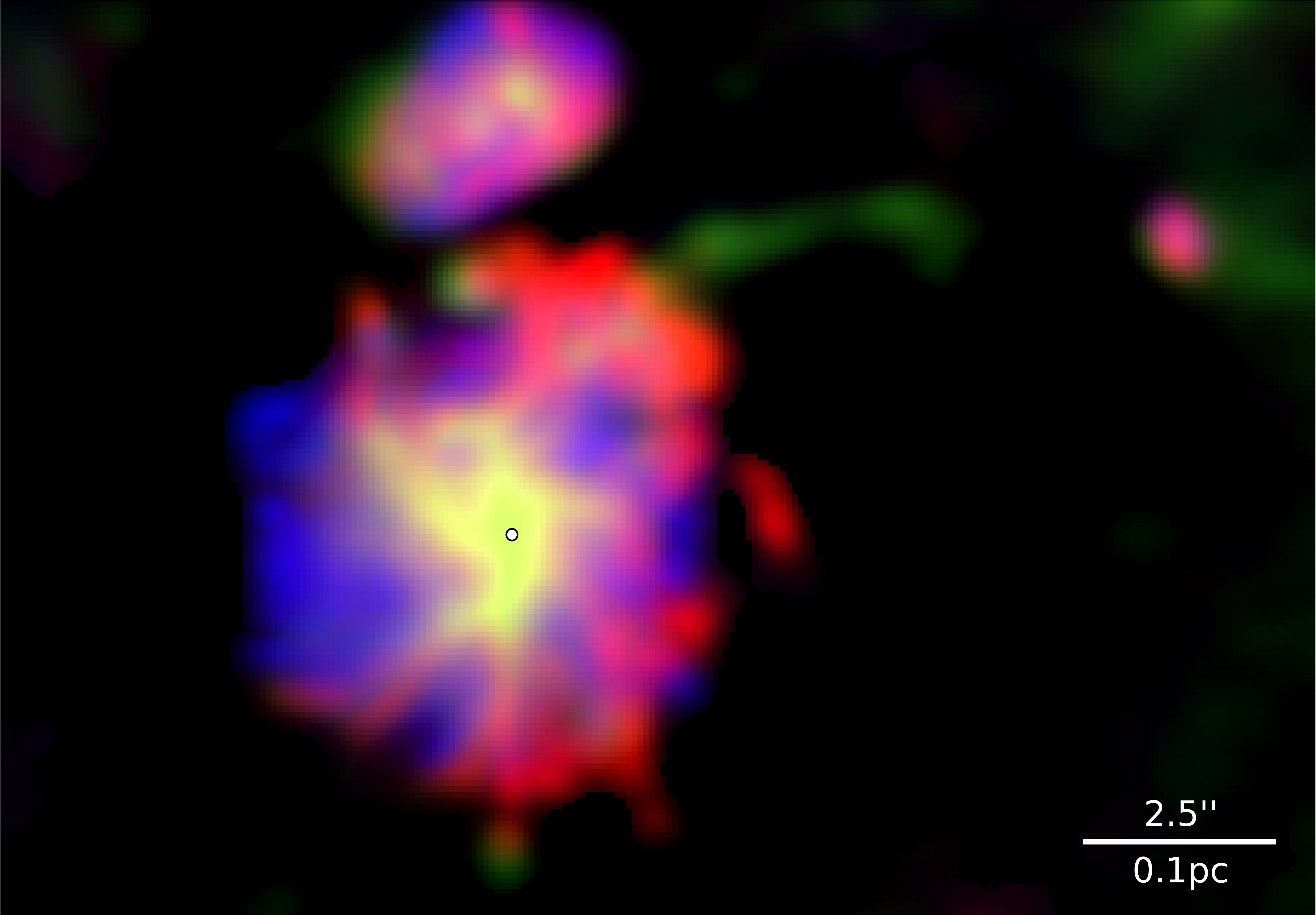}
\caption{Three-color composite image of \SgrB(N). The green image
shows the continuum emission at 242~GHz (Paper II), the red image corresponds
to the molecular species \ce{CH3OCHO}, and the blue image to \ce{C2H5CN}.
The center, which is dominated by the continuum and CH3OCHO emission, appears
yellow. The images of the molecular species have been constructed from stacked
cubes (see more details in Section~\ref{sec:results} and Appendix~A) and
correspond to peak intensity maps. The continuum emission and species like
\ce{CH3OCHO} trace a filamentary structure, while species such as \ce{C2H5CN}
show a spherical or bubble-like shape. The
white circle indicates the position of the central core with coordinates
$\alpha$(J2000)=17$^{h}$47$^{m}$19$^{s}$.87, $\delta$(J2000)=$-$28$^\circ$22$'$18\farcs43.}
\label{fig:rgb_spiral}
\end{figure}
\begin{figure*}
\centering
\includegraphics[width=0.98\textwidth]{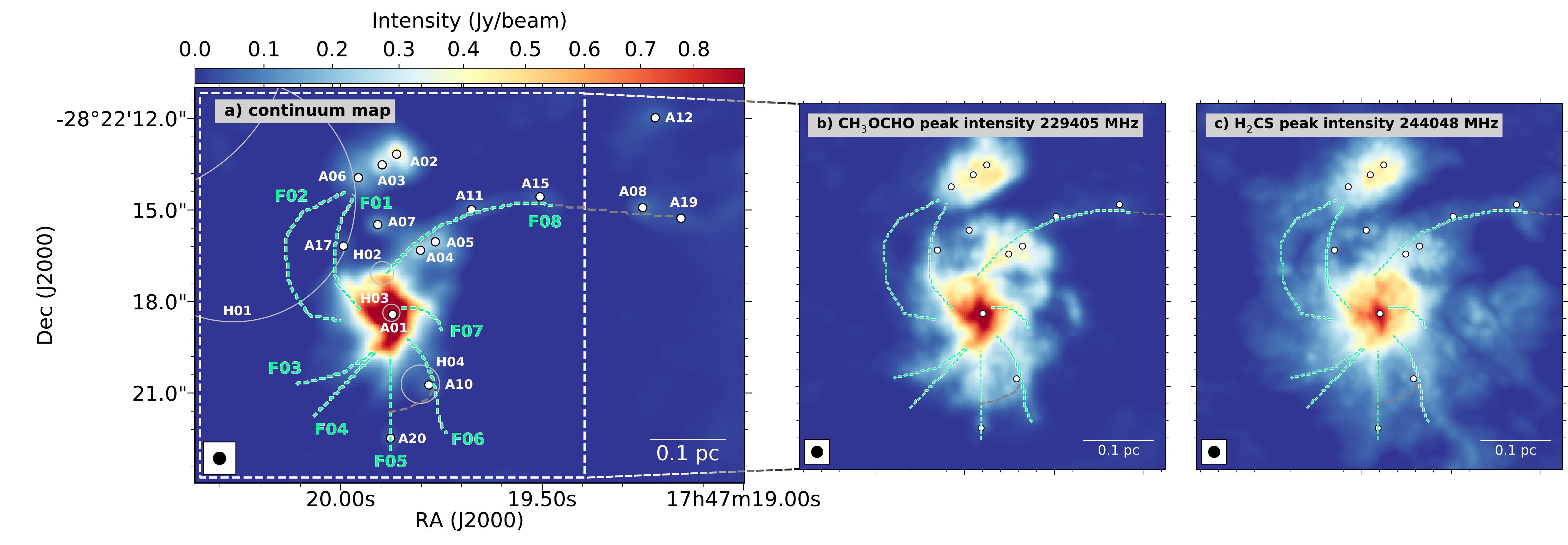}
\caption{\textit{a)} Map of the ALMA 242~GHz continuum emission of
\SgrB(N). The white dots indicate the position of the continuum sources reported
in Paper~II. The green dashed lines trace the path of the filaments identified
in the molecular line data (see Section~\ref{sec: Identification_properties}),
while gray dashed lines trace tentative elongated structures not clearly
confirmed in the molecular emission maps. The light-gray circles indicate
\ion{H}{ii} regions (see \citealt{DePree2014}, and Paper~I). \textit{b)}
Peak intensity map of the bright, isolated transition of \ce{CH3OCHO} at
229.404~GHz. \textit{c)} Peak intensity map of the bright \ce{H2CS} transition
at 244.048~GHz. In all panels, the synthesized beam of $0\farcs4$ is shown
in the bottom-left corner, and a spatial scale bar is shown in the bottom-right
corner.
}
\label{fig:maps_stacked_cubes}
\end{figure*}
The line survey toward \SgrB(N) revealed an extremely rich chemistry in many
of the compact sources reported in Paper~II, with more than 100 lines per
GHz. While the detailed chemistry is the subject of another study (M\"oller
et al.\ in prep), in the current work we focus on species with bright and
clearly identified\footnote{\label{note1}The assignment of transitions to
molecular species is done using the software XCLASS: eXtended CASA Line Analysis
Software Suite: \citet{Moeller2017}, which make uses of the Cologne Database
for Molecular Spectroscopy (CDMS, M\"uller et al.\ 2001, 2005) and Jet Propulsion
Laboratory (JPL, Picket et al.\ 1991) catalogs, via the Virtual Atomic and
Molecular Data Centre (VAMDC, Endres et al.\ 2016).} transitions. Among them,
S-bearing species like \ce{H2CS} and \ce{OCS}, N-bearing species like \ce{CH3CN}
and \ce{C2H5CN}, and O-bearing species like \ce{CH3OH}, \ce{CH3OCHO} and
\ce{CH3OCH3}. We used the most isolated lines of these molecular species
and produced peak intensity maps. These maps reveal two main types of structures
which are illustrated in Fig.~\ref{fig:rgb_spiral}. The dust continuum emission
at 242~GHz (see Paper~II) and species like \ce{CH3OCH3} (dimethyl ether),
\ce{CH3OCHO} (methyl formate), \ce{CH3OH}, \ce{^13CH3OH} (methanol) and \ce{H2CS}
(thioformaldehyde) show a filamentary structure (green and red in Fig.~\ref{fig:rgb_spiral}),
whereas other molecules like \ce{C2H5CN} (ethylcyanid)
or \ce{OCS} (carbonylsulfid) trace instead a spherical or bubble-like shape
(blue in Fig.~\ref{fig:rgb_spiral}). The existence of two different morphologies
in different molecular species hints at opacity effects, excitation conditions
or chemical variations as possible origins. Opacity effects are most likely
to be discarded, because we detect a large number of species
including rare isotopologues (i.e.\ less affected by opacity effects) tracing
both structures. The chemical and excitation properties of both structures
will be discussed in a forthcoming paper (Schwörer et al.\ in prep). Here,
we have focussed on the four molecular species \ce{CH3OCHO}, \ce{CH3OCH3},
\ce{CH3OH,} and \ce{^13CH3OH} included in the spectral-line survey that trace
the filamentary structure and study its kinematic properties.
The analysis of spectral-line surveys with thousands of lines is challenging
because of line blending effects of known and unknown molecular species,
particularly in a chemically-rich region like \SgrB. For any single transition
it is thus not straightforward to accurately determine its spatial distribution
and the kinematic properties (e.g.,\ velocity, linewidth, lineshape). However,
taking advantage of the existence of 100s of lines for each species, we developed
a tool that stacks all transitions of a certain molecule. Line stacking
improves the signal-to-noise ratio, as well as averaging out line blendings.
This permits us to derive the real spatial distribution of the molecular
species, as well as the true line shape and number of (velocity) components.
A more detailed description of the method is presented in Appendix~\ref{append:line_stacking}.
We used the stacked data cubes for different species to characterize the
properties of the filaments in \SgrB(N).
\subsection{Filament identification and properties}
\label{sec: Identification_properties}
The identification of the filaments has been done tracing their peak intensity,
starting at a position close to the dense hub and following the intensity
ridge outwards by visual inspection. Since the central region is optically
thick and most lines appear in absorption against the bright continuum, we
exclude this region. We used peak intensity maps created from stacked cubes
(see Appendix~A), because their higher signal-to-noise ratios allow for a
better determination of faint structures. The identified filaments were also
confirmed in peak intensity maps created from individual molecular transitions.
In total, we identify eight arms or filaments (see Fig. \ref{fig:maps_stacked_cubes})
which are visible in emission of different species. While arms F01, F03,
F04, F05, F06, F07, and F08 emit in the molecules \ce{CH3OCH3}, \ce{CH3OCHO},
\ce{CH3OH} and \ce{^13CH3OH}, arm F02 is only visible in \thiform. Arm F08
is the most extended and seems to connect the central hub with regions located
$\sim0.5$~pc to the west. However, the outer (western) part is mainly visible
in continuum and not in molecular lines which prevents an analysis of its
velocity structure (see Section~\ref{sec:discussion}). This arm leads to
a chalice-shaped area (cf.\ bright red emission in Fig.~\ref{fig:rgb_spiral})
when approaching the central hub, and contains the dense cores A04, A05,
A07 (from Paper~II), and the \ion{H}{ii} region H02 \citep{DePree2014}. On
the contrary, filament F07 appears the shortest, probably due to strong projection
effects. Some filaments (F01, F05, F06, and F08) harbor embedded cores, while
others appear more homogeneous and without clear hints of fragmentation along
them, suggesting the existence of different physical conditions (Schw\"orer
et al.\ in prep). The northern satellite core (containing dense cores A02,
A03, and A06, and being one of the main targets in the search of complex
molecular species, e.g.\ \citealt{Belloche2013}) seems to be connected by
arms F01 and F02 to the main hub (see Section~\ref{sec:discussion} for further
detail). However, arm F01 seems broken, maybe caused by feedback of the neighbor
\ion{H}{ii} region H01. The area to the southwest, in between filaments F05
and F07 looks disordered, and the possibility of another underneath filament
can not be excluded (see gray dashed lines in Fig.~\ref{fig:maps_stacked_cubes},
see also Fig.~\ref{fig:rgb_spiral}).
\begin{figure}
\includegraphics[width=0.95\columnwidth]{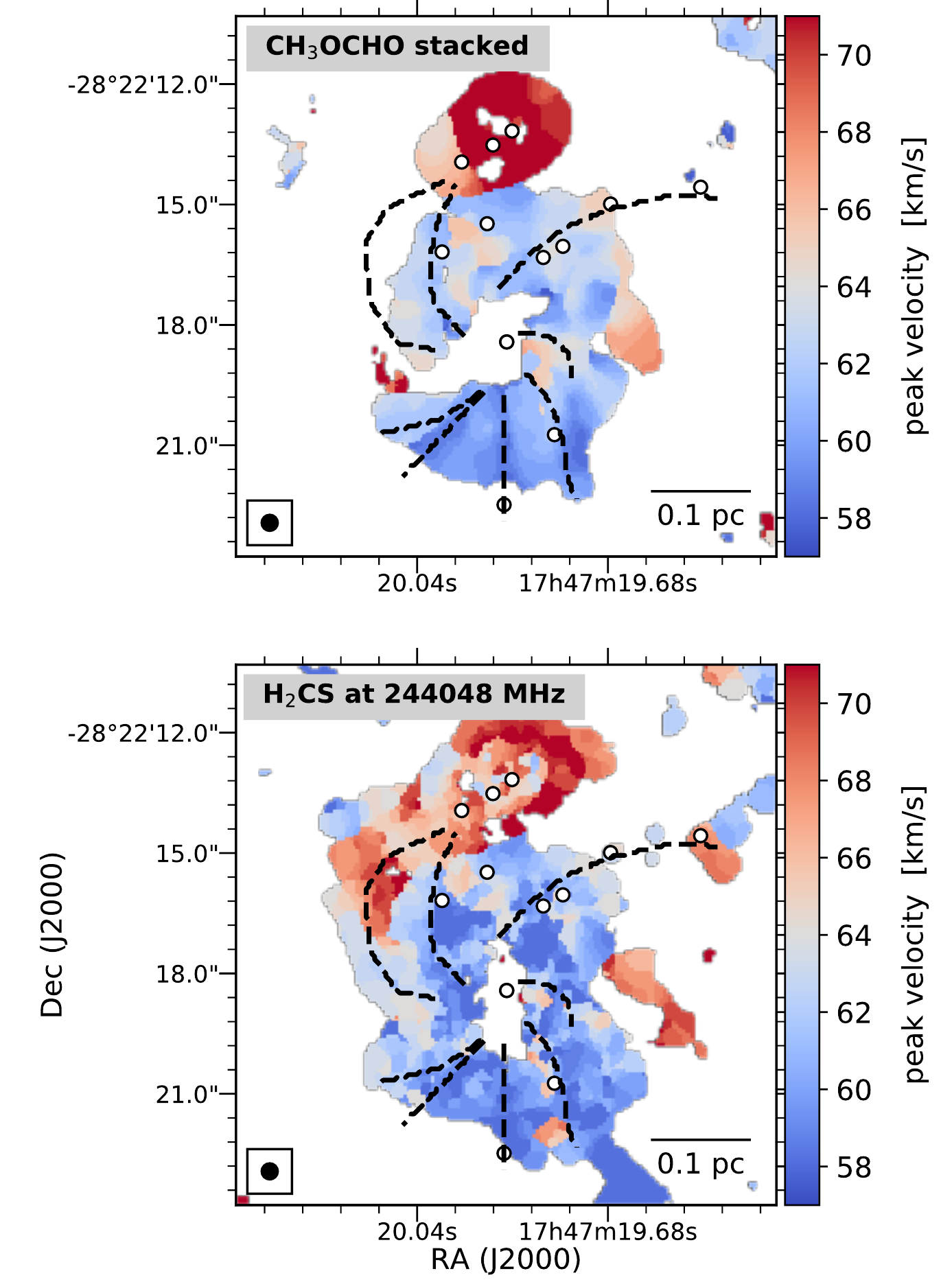}
\caption{Peak velocity map of \ce{H2CS} (bottom-panel) and \ce{CH3OCHO}
(top-panel, using 143 stacked lines, see Appendix A). The emission below
10~$\sigma$ and the center are masked out. The filaments and position of
dense cores are marked with black dashed lines and white dots, respectively.}
\label{fig:velo_CH3OCHO}
\end{figure}
\subsection{Filament kinematics}
\label{sec: filament_kinematics}
\begin{figure}
\centering
\includegraphics[width=0.95\columnwidth]{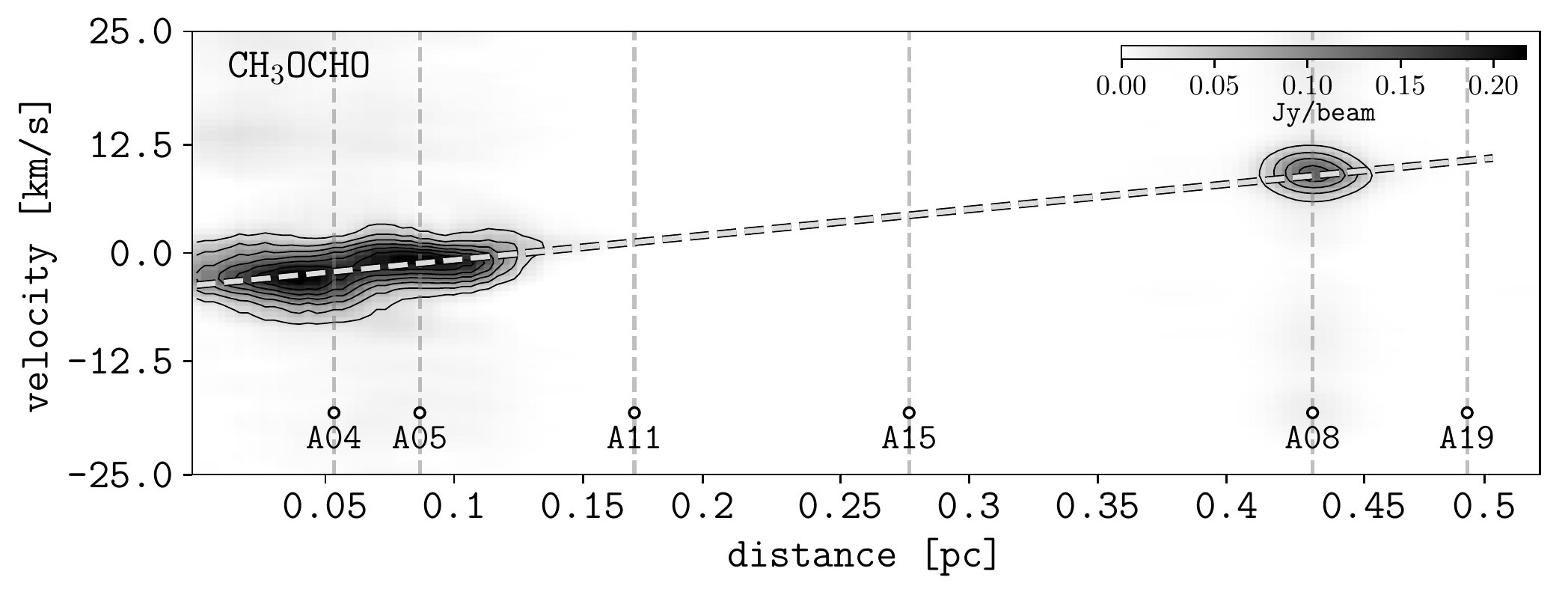}
\includegraphics[width=0.95\columnwidth]{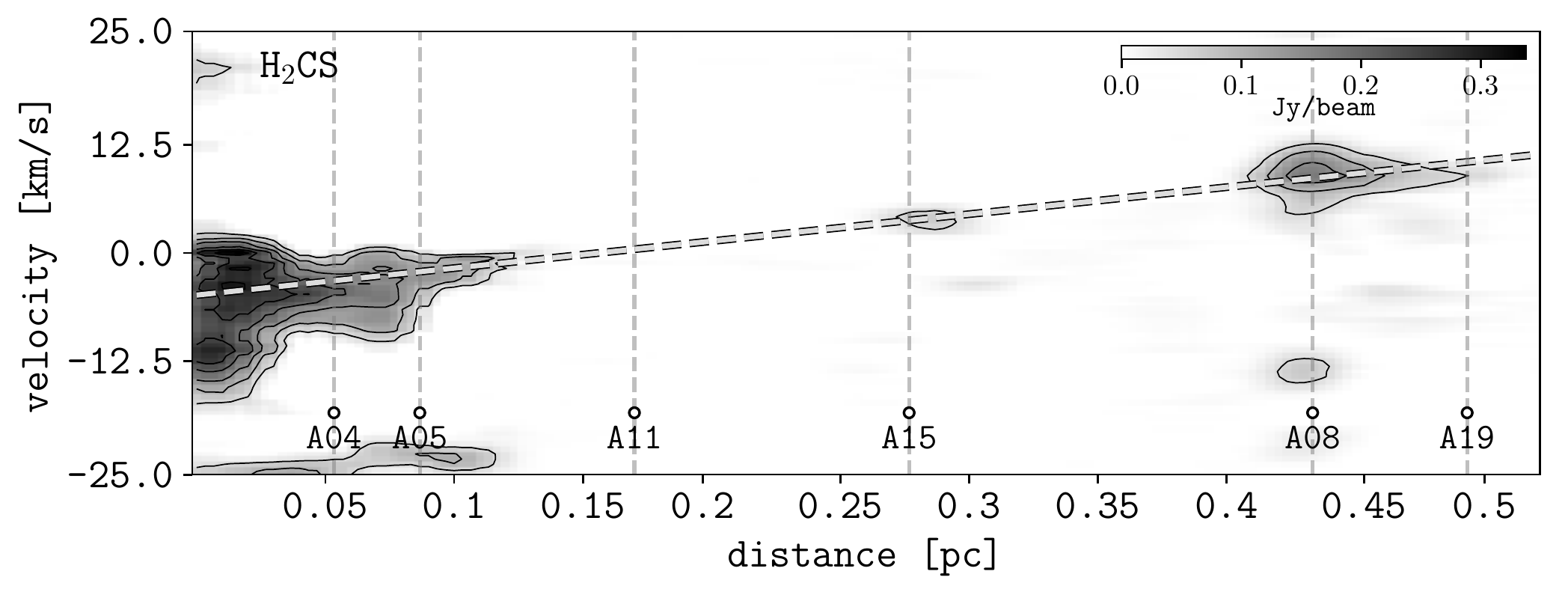}
\caption{Position velocity cut along filament F08 in \ce{CH3OCHO}
and \ce{H2CS}. The emission to the left corresponds to the filament close
to the central hub. The emission at about 0.25~pc and 0.4~pc corresponds
to the cores A15 and A08, A19. The white dashed line corresponds to a velocity
gradient of $\sim$16~km~s$^{-1}$~pc$^{-1}$, consistent with the mean velocity
gradient of F08 (see Table \ref{Table_FreeFall}).} \label{p_v_plot_H2CS_arm_8}
\end{figure}

\begin{figure*}
\centering
\includegraphics[width=1\textwidth]{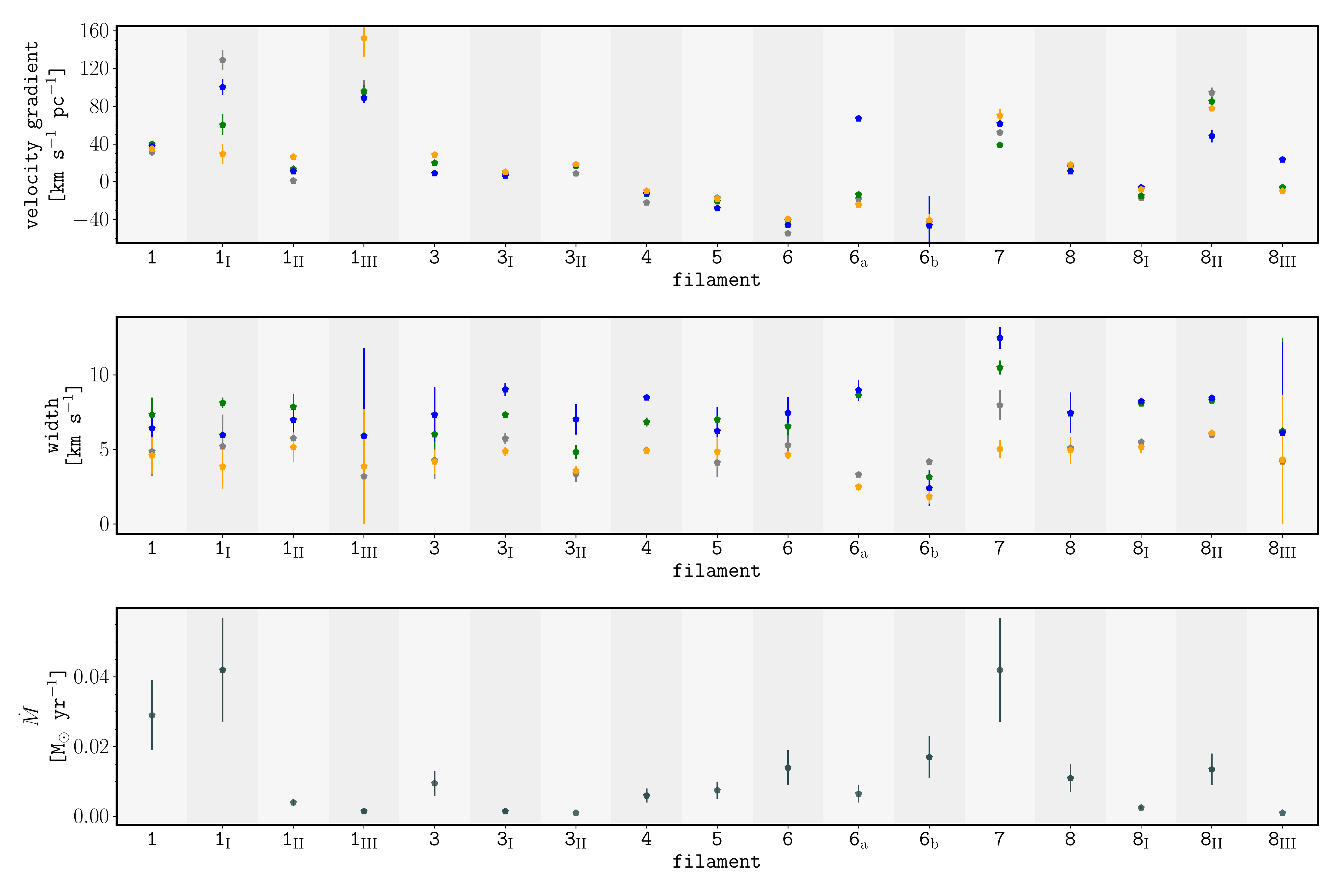}
\caption{Overview picture of the velocity gradients (\textit{top
panel}), velocity linewidths (\textit{middle panel}) and mass accretion rates
(\textit{bottom panel}) along all filaments and subfilaments (indicated by
roman numbers) in \SgrB(N), see Sections~\ref{sec: Identification_properties}
and \ref{sec: filament_kinematics}. Additional velocity components in filament
F06 are marked as 6a and 6b. The gray, green, blue, and orange symbols correspond
to the molecules \ce{CH3OCHO}, \ce{CH3OCH3}, \ce{CH3OH} and \ce{^13CH3OH},
respectively.} \label{fig:velocity_gradient_overview}
\end{figure*}

The analysis of the velocity structure of each filament has been done using
the stacked cubes, excluding arm F02 which is only visible in \ce{H2CS}.
This molecule has, in the frequency range of our line survey, only four transitions,
thus the statistical method of line-stacking offers no significant advantages.
The peak velocity map of the stacked transitions of \ce{CH3OCHO} (see
Fig.~\ref{fig:velo_CH3OCHO})
reveals already a clear velocity gradient from red to blue-shifted velocities
between the northern satellite core and the main hub. Overall, we see a smooth
velocity structure, except at the interface between the northern and main
cores, where several velocity components are present.
A more detailed study of the velocity structure is obtained by producing
position-velocity (pv) cuts along the filaments and by fitting Gaussian functions to the spectra. In Appendix~B
we show the pv-cuts for different molecular species. Most of the filaments
are well described by a single velocity component, with the exception of
filament F06, for which we consider two velocity components. The Gaussian
fits of each spectrum along the filament allow us to derive the velocity
and linewidth of each species, and better explore possible variations with
distance along the filaments (see Appendix~\ref{append:velocity_gradients}).
For all the filaments, we see velocity gradients in all four species, and
we fit them with a linear function. In Table~\ref{tab:filament_parameter},
we list the velocity gradients of seven filaments for three species. The
velocity structure of each filament is consistent between different molecular
species, suggesting that these species trace gas with similar kinematic properties.
We determine an average velocity gradient of roughly 20~km~s$^{-1}$~pc$^{-1}$.
Arm F07 reaches the highest values of 52--70~km~s$^{-1}$~pc$^{-1}$, which
together with its short length, suggests that F07 is a filament oriented
close to the line of sight. While the gradients in filaments F04, F05, and
F06 are negative, meaning that they\ go from red to blue-shifted velocities
when moving from the main hub outwards, the gradients in filament F03 and
F07 are positive. The pv-cut along the most extended filament F08 shows a
connection in its velocity structure between the main hub and the far away
western dense cores A15 located at a distance of $\sim$0.27~pc and A08/A09
located at a distance of $\sim$0.5~pc (see Fig.~\ref{p_v_plot_H2CS_arm_8}).
The kinematic structure of some filaments shows clear variations in the
trend of their velocity structure (see e.g.\ F03, F08, Fig.~\ref{fig:velocity_curve},
and Fig.~\ref{fig:velocity_curve_2}). This is likely due to the filaments
not being perfect straight lines, but having some curvature in the 3D space,
which may result in different velocity gradients (along the filaments) when
seen in projection. We used these variations to define subsections (labeled
with roman numbers) along these filaments (see Appendix~\ref{fig:violin_velocity}).
We divide filament F03 in two sections from 0.02 to 0.055~pc, and 0.07 to
0.12~pc, and filament F08 in three sections from 0.01 to 0.04~pc, 0.055 to
0.07~pc, and 0.08 to 0.11~pc. Filament F01 has a weak additional velocity
component, best visible in \ce{CH3OH} (see Fig.~\ref{fig:PV-Plot}), but also
in \ce{CH3OCHO}, connecting the northern satellite core with the main hub.
We divided F01 into three sections from 0 to 0.03~pc, 0.032 to 0.11~pc, and
0.115 to 0.018~pc. Moreover, as discussed above (see also Fig.~\ref{fig:PV-Plot}),
F06 has two velocity components in the range 0 to 0.04~pc, that we labeled
as F06a and F06b. In addition to the velocity gradient derived for each filament
as a single entity, we calculate the velocity gradient for every identified
segment (see Table~\ref{tab:filament_parameter}). In Fig.~\ref{fig:velocity_gradient_overview}
(top panel), we present an overview of the velocity gradients of each filament
and subfilament.
The variation of the velocity linewidth along the filament for each species
is shown in Fig.~\ref{fig:fwhm_curve}, while the mean linewidths, evaluated
as the average Gaussian width along the filament, are shown in
Fig.~\ref{fig:velocity_gradient_overview}
(middle panel). Typical linewidths range from 3 to 8~km~s$^{-1}$, with a
decreasing trend when moving from the central hub outwards. Along filaments
F03, F04, F05, and F08, the linewidth of \ce{CH3OCHO} and \ce{^13CH3OH} behave
quite similar, while the linewidth of \ce{CH3OH} is in most cases twice the
linewidth of the other species. This is further studied in a forthcoming
paper (Schw\"orer et al.\ in prep). Filament F05 has a local linewidth minimum
at $\sim$0.075~pc, coincident with the position of another filament, which
is not unambiguously identified in our current data (see Fig.~\ref{fig:maps_stacked_cubes}
and Sect.~\ref{sec: Identification_properties}).
\begin{table*}
\caption{Kinematic and physical properties of the filaments in
\SgrB(N)}\label{tab:filament_parameter}
\centering
\begin{tabular}{l c c c c c c c c c}
\hline\hline
\noalign{\smallskip}
{} & \multicolumn{4}{c}{Velocity gradient (km s$^{-1}$
pc$^{-1}$)} && \multicolumn{4}{c}{Filament parameter\tablefootmark{a}} \\
\cline{2-5}\cline{7-10}
\noalign{\smallskip}
{ID} & \ce{CH3OCHO} & \ce{CH3OCH3} &
\ce{CH3OH} & \ce{^13CH3OH} && $M$ ($M_\odot$) & $L$ (pc) & $M/L$
($M_\odot$ pc$^{-1}$) & $\dot{M}$ ($M_\odot$ yr$^{-1}$) \\
\hline
\noalign{\smallskip}
F01 & \phn$+$31.5$\pm$4\phn & \phn$+$40.0$\pm$3\phn &
\phn$+$37.6$\pm$2\phn & \phn$+$34.9$\pm$2\phn && 1050--490 & 0.14 & 7500--3500
& 0.039 -- 0.019\\
--- I & $+$129.0$\pm$10 & \phn$+$60.3$\pm$11 &
$+$100.3$\pm$9\phn & \phn$+$29.6$\pm$11 && \phn690--320 & 0.03 & 23000--10800
& 0.057 -- 0.027\\
--- II & \phnn$+$1.2$\pm$2\phn & \phn$+$13.5$\pm$2\phn &
\phn$+$11.0$\pm$2\phn & \phn$+$26.4$\pm$2\phn && \phn380--180 & 0.08 & 4800--2300
& 0.005 -- 0.003\\
--- III & \phn$+$96.4$\pm$12 & \phn$+$95.3$\pm$7\phn &
\phn$+$88.9$\pm$6\phn & $+$152.2$\pm$20 && 16--7 & 0.03 & 540--250
& 0.002 -- 0.001\\ \hline \noalign{\smallskip}
F03 & \phn$+$20.0$\pm$1\phn & \phn$+$19.9$\pm$2\phn &
\phnn$+$9.0$\pm$3\phn & \phn$+$28.6$\pm$2\phn && \phn620--290 & 0.13 & 4800--2200
& 0.013 -- 0.006\\
--- I & \phnn$+$8.4$\pm$3\phn & \phnn$+$6.8$\pm$2\phn &
\phnn$+$6.5$\pm$1\phn & \phn$+$10.4$\pm$2\phn && 200--95 & 0.03 & 6800--3200
& 0.002 -- 0.001\\
--- II & \phnn$+$8.8$\pm$2\phn & \phn$+$16.8$\pm$1\phn &
\phn$+$17.7$\pm$1\phn & \phn$+$18.5$\pm$1\phn && \phn40--18 & 0.05 & 800--380
& 0.001 -- 0.001\\ \hline \noalign{\smallskip}
F04 & \phn$-$22.1$\pm$2\phn & \phn$-$12.8$\pm$2\phn &
\phn$-$12.4$\pm$2\phn & \phnn$-$9.7$\pm$1\phn && \phn570--270 & 0.14 & 4100--1900
& 0.008 -- 0.004\\ \hline \noalign{\smallskip}
F05 & \phn$-$17.0$\pm$2\phn & \phn$-$20.7$\pm$1\phn &
\phn$-$28.0$\pm$1\phn & \phn$-$18.1$\pm$1\phn && \phn490--230 & 0.13 & 3800--1800
& 0.010 -- 0.005\\ \hline \noalign{\smallskip}
F06 & \phn$-$54.7$\pm$2\phn & \phn$-$41.0$\pm$2\phn &
\phn$-$46.0$\pm$2\phn & \phn$-$39.4$\pm$2\phn && \phn410--190 & 0.16 & 2500--1200
& 0.019 -- 0.009\\
--- a & \phn$-$18.1$\pm$2\phn & \phn$-$13.6$\pm$1\phn &
\phn$+$67.1$\pm$1\phn & \phn$-$24.3$\pm$1\phn && \phn500--240 & 0.02 & 25000--12000
& 0.009 -- 0.004\\
--- b & \phn$-$43.1$\pm$2\phn & $-$103.3$\pm$17 &
\phn$-$46.4$\pm$31 & \phn$-$40.7$\pm$7\phn && \phn500--240 & 0.02 & 25000--12000
& 0.023 -- 0.011\\ \hline \noalign{\smallskip}
F07 & \phn$+$52.3$\pm$4\phn & \phn$+$38.9$\pm$3\phn &
\phn$+$61.5$\pm$2\phn & \phn$+$70.2$\pm$7\phn && \phn970--460 & 0.10 & 9700--4600
& 0.057 -- 0.027\\ \hline \noalign{\smallskip}
F08 & \phn$+$18.0$\pm$2\phn & \phn$+$16.1$\pm$2\phn &
\phn$+$11.0$\pm$1\phn & \phn$+$18.2$\pm$2\phn && \phn810--380 & 0.30 & 2700--1300
& 0.015 -- 0.007\\
--- I & \phn$-$17.3$\pm$2\phn & \phn$-$14.9$\pm$1\phn &
\phnn$-$6.1$\pm$4\phn & \phnn$-$8.1$\pm$1\phn && 200--95 & 0.03 & 6700--3200
& 0.003 -- 0.002\\
--- II & \phn$+$94.5$\pm$6\phn & \phn$+$85.2$\pm$5\phn &
\phn$+$48.5$\pm$7\phn & \phn$+$77.7$\pm$3\phn && \phn210--100 & 0.01 & 21000--9900\phn
& 0.018 -- 0.009\\
--- III & \phnn$+$5.8$\pm$4\phn & \phnn$-$6.2$\pm$3\phn &
\phn$+$23.5$\pm$1\phn & \phnn$-$9.9$\pm$2\phn && 150--70 & 0.03 & 5100--2400
& 0.001 -- 0.001\\
\hline
\end{tabular}
\tablefoot{
\tablefoottext{a}{Mass accretion rates have been computed
with the median of the velocity gradients of the four species.}
}
\end{table*}
\begin{figure*}
\centering
\includegraphics[width=1\textwidth]{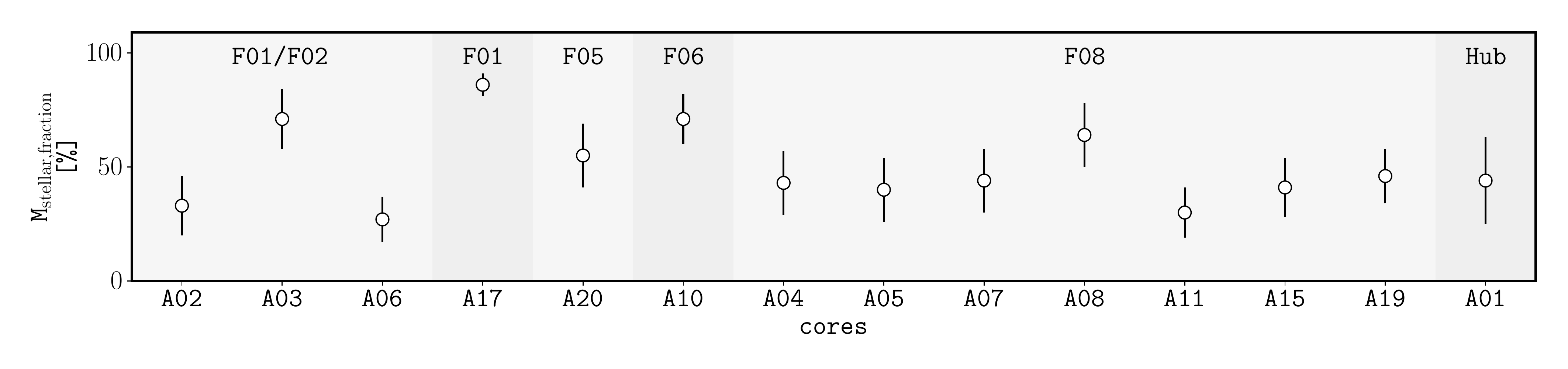}
\caption{Stellar mass fraction (see Sect.~\ref{sec:core_properties_accretion_time_scale})
of the dense cores. Cores are grouped by their host filaments.}
\label{fig:overview_stellar_mass_ratio}
\end{figure*}
\begin{figure}
\centering
\includegraphics[width=0.95\columnwidth]{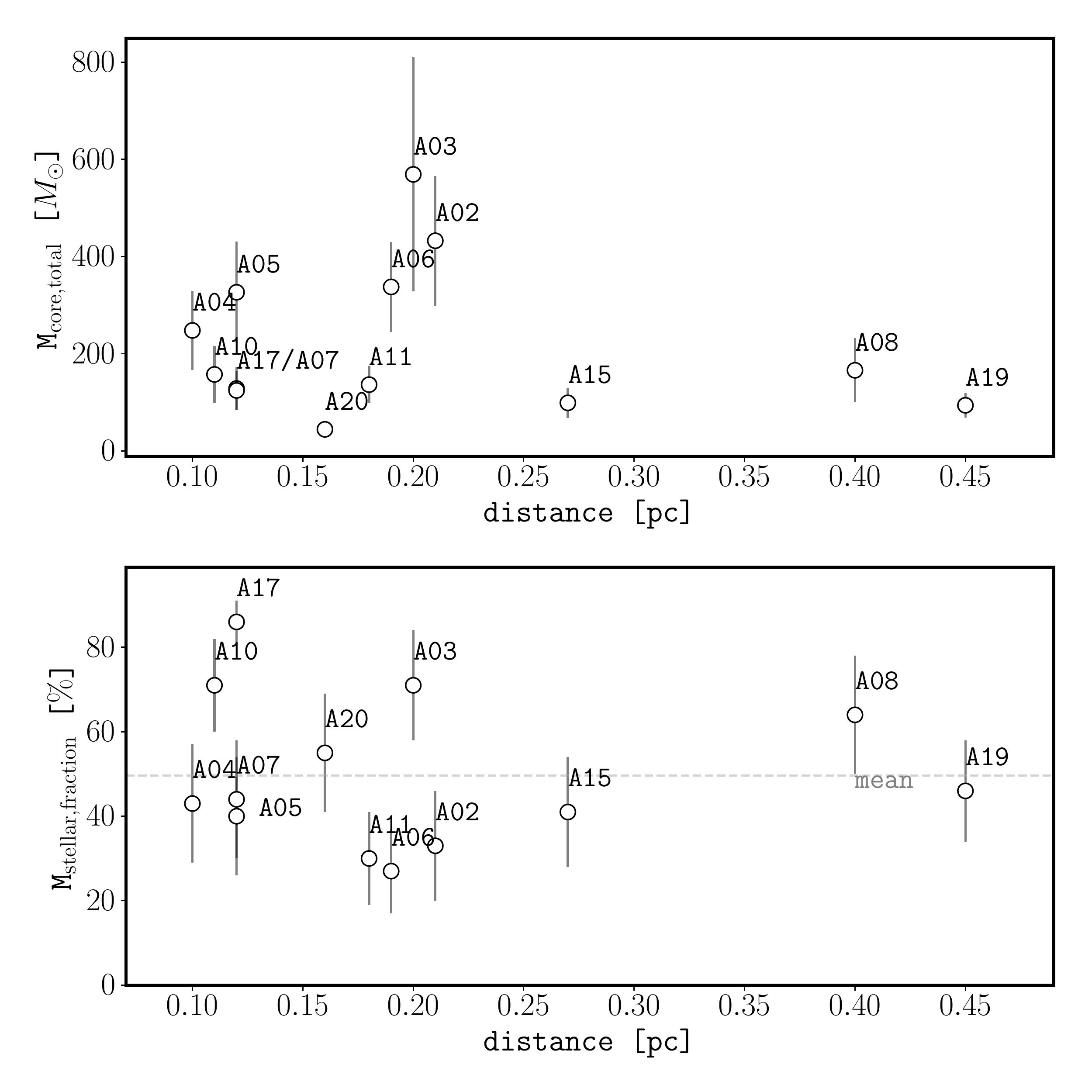}
\caption{\textit{Top panel:} Total mass of the dense cores against
their distance to the main hub. \textit{Bottom panel:} Stellar mass fraction
of the dense cores against their distance to the main hub. The gray dashed
line indicates the mean stellar mass fraction of of 50\%.} \label{fig:M_core_dist}
\end{figure}
\begin{table*}
\caption{Properties and timescales of the dense cores located within
filaments in \SgrB(N)} \label{Table_FreeFall}
%\sisetup{round-mode=places, round-precision=2}
\centering
\begin{tabular}{rcccccccccc}
\hline\hline
\noalign{\smallskip}
\multicolumn{1}{c}{} & $\mathrm{\theta_{core}}$\tablefootmark{a}
& $d_\mathrm{hub}$\tablefootmark{b} & $T$\tablefootmark{c} & $L$ &
$M_\mathrm{d+g}$ & $M_\mathrm{stellar}$\tablefootmark{d} & $M_\mathrm{stellar,ratio}$
& $t_\mathrm{core,ff}$ & $t_\mathrm{fil,ff}$ \\
\multicolumn{1}{c}{ID} & (\arcsec) & (pc) & (K) &
($\times$10$^3$ $L_\odot$) & ($M_\odot$) & ($M_\odot$) & (\%) & (kyr)
& (kyr) \\
\hline
\noalign{\smallskip}
F01~A02\tablefootmark{e} & 0.64 & 0.21 & 120\phn & 61 & \phn300 $\pm$ 45\phn
& \phn90 -- 220 &
20 -- 45 & \phn8 $\pm$ 0.3 & 10 -- 16 \\
A03\tablefootmark{e} & 0.70 & 0.20 & 195\phn & 510 & \phn140 $\pm$ 20\phn
& 230 -- 650 &
60 -- 85 & 13 $\pm$ 0.6 & \phn9 -- 15 \\
A06\tablefootmark{e} & 0.90 & 0.19 & 80\phn\phn & 24 & \phn255 $\pm$ 35\phn
& \phn60 -- 140 &
20 -- 40 & 14 $\pm$ 0.6 & \phn9 -- 13 \\
A17\phnx & 0.53 & 0.12 & 110\phn & 29 & \phn\phn15 $\pm$ 1\phn\phn
&
\phn70 -- 160 & 80 -- 90 & 27 $\pm$ 1.0 &
4 -- 7 \\
\midrule
F05~A20\phnx & 0.44 & 0.16 & 50\tablefootmark{f}\phnx & 0.87 & \phn\phn20
$\pm$ 5\phn\phn & 20 -- 35 & 40 -- 70
& 18 $\pm$ 0.9 & \phn7 -- 11 \\
\midrule
F06~A10\phnx & 0.64 & 0.11 & 105\phn & 36 & \phn\phn40 $\pm$ 5\phn\phn
&
\phn70 -- 170 & 60 -- 80 & 22 $\pm$ 0.9 &
4 -- 6 \\
\midrule
F08~A04\phnx & 0.77 & 0.10 & 170\phn & 360 & \phn140 $\pm$ 20\phn
&
\phn70 -- 170 & 30 --60 & 15 $\pm$ 0.6 &
3 -- 5 \\
A05\phnx & 0.84 & 0.12 & 100\phn & 51 & \phn200 $\pm$ 30\phn
&
\phn80 -- 200 & 25 -- 55 & 15 $\pm$ 0.6 &
4 -- 7 \\
A07\phnx & 0.44 & 0.12 & 85\phn\phn & 7.3 & \phn\phn70 $\pm$ 10\phn
&
35 -- 85 & 30 -- 60 & \phn9 $\pm$
0.4 & 4 -- 7 \\
A08\phnx & 0.56 & 0.40 & 110\phn & 33 & \phn\phn55 $\pm$ 10\phn
&
\phn65 -- 170 & 50 -- 80 & 15 $\pm$ 0.7 &
27 -- 42 \\
A11\phnx & 0.84 & 0.18 & 50\tablefootmark{f}\phnx & 3.2 & \phn100 $\pm$
15\phn & 30 -- 60 &
20 -- 40 & 21 $\pm$ 0.9 & \phn8 -- 13 \\
A15\phnx & 0.80 & 0.27 & 50\tablefootmark{f}\phnx & 2.9 & \phn\phn60 $\pm$
10\phn & 30 -- 60 & 30 -- 55
& 25 $\pm$ 1.1 & 15 -- 23 \\
A19\phnx & 0.83 & 0.45 & 50\tablefootmark{f}\phnx & 3.1 & \phn\phn50 $\pm$
5\phn\phn & 30 -- 65 &
35 -- 60 & 29 $\pm$ 1.1 & 32 -- 50 \\
\midrule
Hub~A01\phnx & 1.45 & -- & 250\tablefootmark{f} & 5900 & 1960 $\pm$ 280
& \phn800 -- 3000 &
25 -- 65 & 11 $\pm$ 0.4 & -- \\
\hline
\noalign{\smallskip}
\end{tabular}
\tablefoot{
\tablefoottext{a}{Angular radius ($\theta_{core}$) of the
cores as listed in Paper II}
\tablefoottext{b}{The distance between the core and the hub
($d_\mathrm{hub}$) is determined from a straight line connecting both, and
can, therefore, be a lower limit to the real distance when considering projeciton
effects.}
\tablefoottext{c}{The temperatures ($T$) have been determined
for different molecular species (M\"oller et al.\ in prep) resulting in an
error of 10\%. After propagation of these errors, we have determined the
uncertainties in the gas mass and timescales.}
\tablefoottext{d}{The lower value of the stellar mass is
obtained from the luminosity listed in the table after decreasing it by a
factor of eight. This aims to reproduce: (i) a core radius two times smaller
than the measured size, and (ii) a contribution of 50\% to the total luminosity
coming from accretion and not from stellar radiation.}
\tablefoottext{e}{Cores A02, A03, and A06 can also be connected
to the central hub via filament F02.}
\tablefoottext{f}{Temperatures are assumed, rather than obtained
by fitting.}}
\end{table*}
\subsection{Mass accretion rates and filament stability}
\label{sec: mass_accretion}
We evaluated the mass accretion rate of each filament using its mass and
the derived velocity gradients. The mass of each filament was
estimated from the dust continuum emission at 242~GHz (see Paper~II).
For optically thin emission, a dust opacity of 0.899~cm$^2$/g \citep{Ossenkopf1994},
a
dust-to-gas mass ratio of 100, and dust temperatures between 50--100~K,
we derive masses of the filaments in the range 200--1000~$M_\odot$
(see Table~\ref{tab:filament_parameter}). The higher mass values for
filaments F01 and F08 are due to the elongation of F01 into the central
hub and the presence of the bright chalice-shaped structure surrounding
F08. Since filaments F03 and F04 are located close together in the plane
of the sky, part of their mass might be double-counted, although the effect
should be negligible, since the overlap of the filaments occurs only in about
one fourth of their extend.
Following \citet{Kirk2013}, we calculate the mass accretion rate implied
by our velocity gradients assuming a cylindrical model for our filaments.
The mass accretion rate ($\dot{M}$) is derived as
\begin{equation}
\dot{M} = v_{||} \cdot \frac{M_\mathrm{fil}}{L_\mathrm{fil}} = \nabla v_{||}
\cdot M_\mathrm{fil},
\end{equation}
where $M_\mathrm{fil}$ is the mass of the filament, $L_\mathrm{fil}$ the
length of the filament and $v_{||}$ the velocity parallel to the filament.
Considering an inclination $\alpha$ between the filament and the plane of
the sky, the parameters that we observe are given by
\begin{equation}
L_\mathrm{fil, obs} = L_\mathrm{fil} \cos(\alpha) \quad\text{and}\quad v_\mathrm{||,
obs} = v_{||} \sin(\alpha),
\end{equation}
which modify the mass accretion rate to
\begin{equation}
\dot{M} = \frac{\nabla v_{||,obs} \cdot M_\mathrm{fil}}{\tan(\alpha)}.
\end{equation}
For the calculation of the mass accretion rates, we assumed a projection
angle $\alpha$ of 45$^\circ$. An angle of 25$^\circ$ roughly doubles our
results, while an angle of 65$^\circ$ halves them. For the velocity gradient,
we used the median of the values determined for all the species (see
Table~\ref{tab:filament_parameter}).
In Fig.~\ref{fig:velocity_gradient_overview} (bottom panel), we plot the
mass accretion rates for all the filaments and subfilaments. The mass accretion
rates along the filaments of \SgrB(N) are between 0.004--0.04~M$_\odot$~yr$^{-1}$
at the scales of 0.1--0.2~pc (see Table \ref{tab:filament_parameter}), thus
$10$--$100$ times larger than rates usually found in star-forming filaments
at larger scales of $\sim$1~pc (e.g., \citealt{Peretto2013}, \citealt{Lu2018},
Trevi\~no-Morales et al.\ in prep). Altogether, considering the average accretion
rates listed in Table~\ref{tab:filament_parameter}, the filaments in \SgrB(N)
accrete at a rate of 0.08--0.16~$M_\odot$~yr$^{-1}$. This results in a total
of 80--160~$M_\odot$ accreted onto the dense hub in about 1000~yr, suggesting
a timescale for the formation of the hub of about 60--300~kyr, assuming current
accretion rates.
In Section~\ref{sec: Identification_properties}, we show that some filaments
appear fragmented and harbor embedded cores, which raises the question of
how stable the filaments are. The mass-to-length ratio ($M/L$) is a measure
of the stability of the filament (e.g., \citealt{Ostriker1964}, \citealt{Fischera2012}).
If this value exceeds a critical limit, the filament will gravitationally
collapse under its weight perpendicular to its main axis. Assuming that turbulence
can stabilize the filament, the critical value is calculated as
\begin{equation}
\left(\frac{M_\mathrm{fil}}{L_\mathrm{fil}}\right)_\mathrm{crit} \approx
\frac{2\sigma^2}{G}
\end{equation}
where $\sigma$ is the velocity dispersion and $G$ the gravitational constant.
Since the measured linewidths are in the range of 3 to 8~km~s$^{-1}$ (see
Fig.~\ref{fig:velocity_gradient_overview}), we can neglect the thermal contribution,
which corresponds to a few 0.1~km~s$^{-1}$ for the considered species at
$50$--$100$~K. This results in a value of $(M/L)_\mathrm{crit}$ about ($4$--
$30)\times10^3$~$M_\odot$~pc$^{-1}$.
According to our measured mass-to-length ratio (see Table~\ref{tab:filament_parameter}),
the filaments in \SgrB(N) seem to be stable under gravitational collapse.
The variation of mass-to-length ratio along the filaments by assuming different
temperatures is presented in Fig.~\ref{fig:M/L_variation} and show a decreasing
trend when moving from the central hub outwards.
\subsection{Dense core properties and accretion timescales}
\label{sec:core_properties_accretion_time_scale}
We determined the temperatures of the dense cores (from Paper~II) located
within the filaments by fitting the spectral lines of different molecular
species with the software XCLASS (Möller et al. in prep). All values are
in the range of 50--200~K and listed in Table~\ref{Table_FreeFall}. These
high temperatures suggest the presence of already formed stars inside the
cores\footnote{Other mechanisms such as external or shock heating can be
excluded. For a typical density of $10^6$~cm$^{-3}$ and temperature of 100~K,
a heating source would need to see the core as optically thin around the
peak ($\sim30$~$\mu$m) of the black body emission. The dust absorption coefficient
$\kappa_ {\nu}$ at $30$~$\mu$m is about 300~cm$^2$g$^{-1}$, which yields
a distance of only 50~au. Therefore, the heating source has to be effectively
embedded in order to heat the gas to the measured temperatures. Heating through
shocks is also unlikely since for infall rates of $\leq0.16~M_\odot$ yr$^{-1}$
and a assumed accretion shock velocity of $\sim 8$~km~s$^{-1}$, the derived
luminosities are only $1700$~L$_\odot$, less than the average luminosities
derived for our cores. }.
The stellar content can be derived from the luminosity of the dense cores.
Considering them as spherical black bodies, the stellar luminosity is given
by the Stefan-Boltzmann equation
\begin{equation}
L = 4 \pi r^2 \sigma T^4 \label{eq:stefan-boltzmann-law}
\end{equation}
where $r$ is the radius of the dense cores, $\sigma$ the Stefan-Boltzmann
constant and $T$ the temperature. We present more details regarding the black
body assumption and the validity of the Stefan-Boltzmann law in
Appendix~\ref{append:Black_body}.
The derived luminosities are listed in Table 2. The large gas masses of the
dense cores (see Paper~II and Table \ref{Table_FreeFall}) suggest that not
one single star, but a stellar cluster is forming in each core. While the
relation between stellar mass and luminosity is known for single stars (e.g.,
\citealt{Eker2018}), there is no direct relation for star clusters. Thus,
we have simulated $10^5$ clusters, and determined the final stellar luminosity
and mass (see Appendix \ref{append:M_Lum_Lum} for details). We find that
the cluster luminosity and stellar mass of a cluster follow the relation
\begin{equation}
\mathrm{log} \: \left( \frac{M/M_\odot}{L/L_\odot} \right) = -0.6 \times
\mathrm{log} \left( \frac{L}{L_\odot} \right) + 0.5.
\end{equation}
The stellar mass content, M$_{stellar}$, for each core within the filaments
is listed in Table 2. With this, we calculated the ratio between the stellar
mass and total mass as
\begin{equation}
M_\mathrm{stellar,fraction} = \frac{M_\mathrm{stellar}}{M_\mathrm{d+g}+M_\mathrm{stellar}}.
\end{equation}
All values are summarized in Table \ref{Table_FreeFall} and visualized in
Fig.~\ref{fig:overview_stellar_mass_ratio}.
For all cores, the stellar mass is within the range 20-90\% of the total
mass, with a mean (median) value of 50\% (44\%). We have also investigated
the relation of the mass of the different cores with respect to their distance
to the central hub. In Fig.~\ref{fig:M_core_dist}, we plot the core mass
(top panel) and the stellar mass ratio (bottom panel) as a function of the
distance to the hub. While no striking correlation is found, we see a bimodal
distribution. Cores with masses above 200~$M_\odot$ are located preferentially
at distances $\textless$0.25~pc, while less massive cores can be found up
to distances of about 0.5 pc. Similarly, cores located closer to the central
hub can have a higher stellar mass fraction than those located farther away.
In Sect.~\ref{sec:cluster_formation} we discuss these results in the context
of converging filaments and the formation of a cluster in \SgrB(N).
One further question of interest is whether the dense cores will be accreted
onto the central hub before they are disrupted by internal star formation.
The timescale for forming stars can be estimated by assuming that the dense
cores will collapse on a free-fall timescale
\begin{equation}
t_\mathrm{ff} = \sqrt{\frac{4\pi^2r^3}{32GM_{d+g}}} \label{eq:t_ff}
\end{equation}
where $M$ is the mass of the dense core and $r$ its radius (values reported
in Paper~II). The $t_\mathrm{core,ff}$ values are listed in the last column
of Table~\ref{Table_FreeFall}, and range from $10^3$ to $10^4$~yr .
We compared the $t_\mathrm{core,ff}$ with the time required by the dense
cores to travel from their position to the central dense hub assuming a free-fall
scenario along the filaments, $t_\mathrm{fil,ff}$ (see Table~\ref{Table_FreeFall}).
To estimate the gravitational acting central mass (see Equation~\ref{eq:t_ff}),
we integrated the flux density over the entire dense hub and assume an averaged
temperature of $50$--$100$~K that result, in a mass between ($25$--
$10)\times10^3$~$M_\odot$.
The distance is given by the separation between the core and the central
hub. The $t_\mathrm{fil,ff}$ ranges from $10^ 3$ to $10^4$~yr, comparable
to the timescale over which a core collapses and forms a stellar cluster.
%--------------------------------------------------------------------
\section{Discussion}\label{sec:discussion}
\subsection{Converging filaments}
\begin{figure}
\centering
\includegraphics[width=0.98\columnwidth]{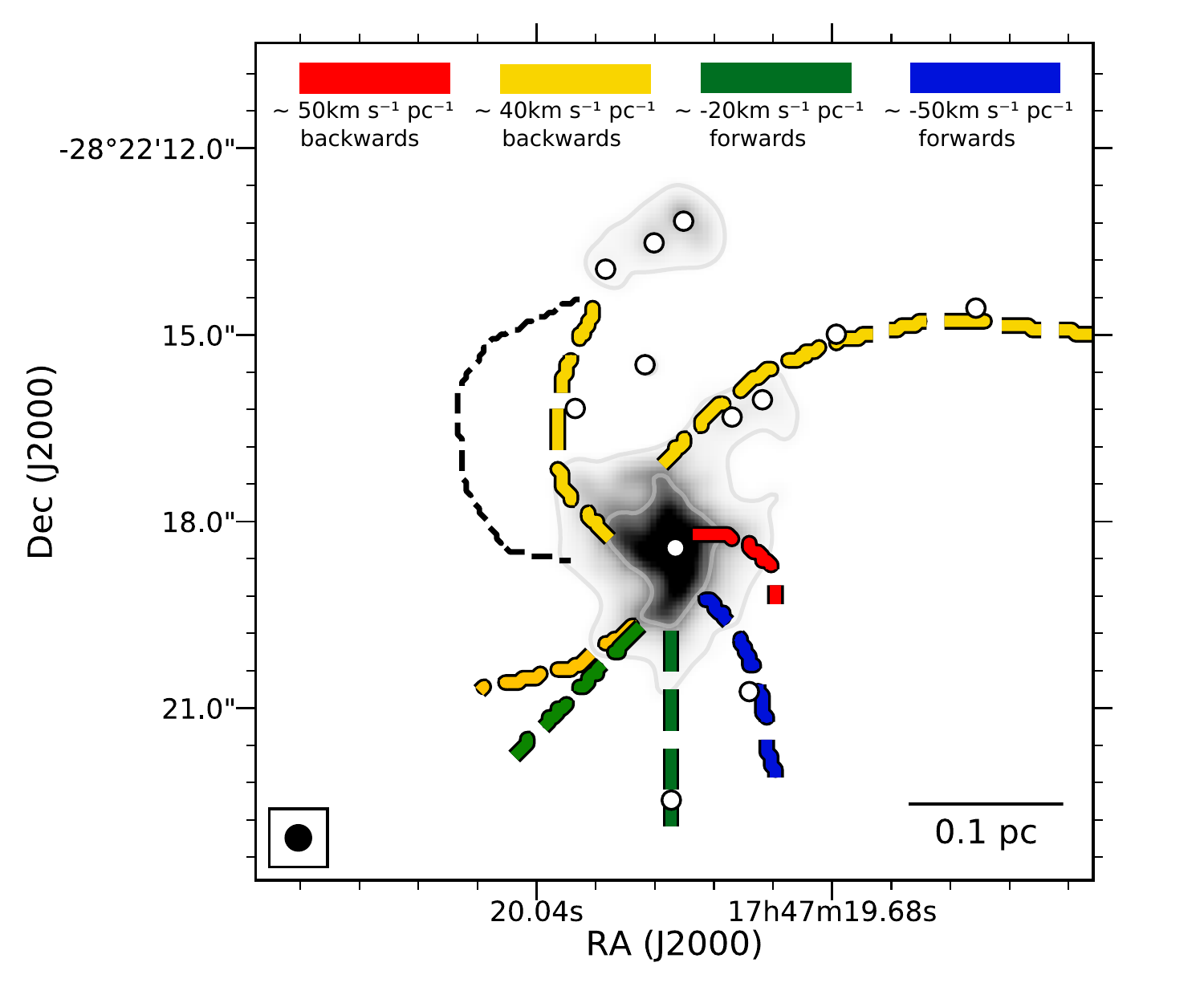}
\caption{Filaments in \SgrB(N) colored by their mean velocity gradient
as derived for the molecular species \ce{CH3OCHO}, \ce{CH3OCH3}, \ce{CH3OH}
and \ce{^13CH3OH}. The velocity gradient ranges from $-$50~km~s$^{-1}$~pc$^{-1}$
(blue), corresponding to filaments located in front, to $+$50~km~s$^{-1}$~pc$^{-1}$
(red), corresponding to filaments located in the back. The gray contours
indicating 242~GHz continuum emission at a level of 1.2~Jy~beam$^{-1}$. The
white circles represent the dense cores, and the black dashed line shows
the filament F02, only visible in \ce{H2CS}.} \label{fig:blank_filaments}
\end{figure}

The analysis of transitions of different molecular species revealed a network
of filaments in \SgrB(N), converging toward a central massive core. We derived
velocity gradients along the filaments, which are 10--100 times larger than
is usually found in other star-forming regions at larger scales (e.g.\ \citealt{Peretto2013},
\citealt{Lu2018}). We have considered here if these large velocity gradients
can be caused by other mechanisms rather than accretion to the center, for
example expanding motions associated with outflows or explosive events like
the one seen in Orion~KL (e.g.\ Bally et al.\ 2017). For outflows, typical
linewidths are 30--100~km~s$^{-1}$ (e.g.\ Wu et al.\ 2004, \citealt{Lopez-Sepulcre2010},
\citealt{Sanchez2013}), roughly ten times larger than what we measure. Furthermore,
common outflow tracers are \ce{CO}, \ce{HCO+} and \ce{SiO} (e.g.\ Wu et al.\
2004, \citealt{Sanchez2013}, but see also \citealt{Palau2011}, 2017) and
not complex molecules like \ce{CH3OCHO} or \ce{CH3OCH3} which we find tracing
the filamentary structure of \SgrB(N). Explosion events such as that in Orion~KL
are also a doubtful explanation since they produce clear and straightforward
radial structures. Contrary to this, the structures that we identified in
\SgrB(N) appear curved and bent, which leaves us with accreting filaments
as the most plausible explanation.
For cylindrical, spatially tilted filaments, the transport of material
along them toward the dense hub (i.e.\ velocities close to the center
are higher than outwards) would result in different signs of the
velocity gradient, as we find in \SgrB(N) (see
Fig.~\ref{fig:velocity_gradient_overview} and
Table~\ref{tab:filament_parameter}). In this scenario, positive
velocity gradients correspond to filaments extending backwards, and
vice-versa, allowing us to have an idea of the possible 3D structure of
the region. In Fig.~\ref{fig:blank_filaments} we show the distribution
of filaments in \SgrB(N) colored by velocity gradient, in which red
corresponds to positive velocity gradients (filaments located in the
back) and blue to negative velocity gradients
(filaments located in front). A remarkable case in \SgrB(N) are the
neighboring filaments F03 and F04. Filament F03 would be located in the
back while F04 in front of the hub. Overall, filaments F03, F07, and F08
would be located in the back, and F04, F05, and F06 would be in front
(see Fig.~\ref{fig:blank_filaments}). In addition to the velocity field,
we find that the linewidths for many filaments (e.g.\ F01, F03, F05, F06,
and F08, see Fig.~\ref{fig:fwhm_curve}) increase when approaching the center\footnote{We
note that the increase of linewidth toward the center may be partially caused
by the line stacking method. The positions along the filament close to the
center show a richer chemistry (i.e., more line features). This excess of
lines, compared to the outer positions, may result in a artificial increase
of the linewidth (see Appendix \ref{append:line_stacking} for further detail).}.
This effect, combined with the increasing mass-to-length ratio (derived
independently from the continuum emission) when moving inwards, suggests
that the accretion or infall rates onto the filaments may be increasing in
the vicinity of the main hub.
Finally, some filaments appear extended, connecting multiple dense
cores with the main hub. In particular, filament F08 extends up to the
distant western cores A08 and A19, located at distances $\sim$0.5~pc.
Although the molecular line emission is not bright in the filament
between core A05 and cores A08 and A19, the velocity in A19 fits the
velocity gradient obtained in the region of the filament closer to the
hub (see Fig. \ref{p_v_plot_H2CS_arm_8}). This suggests that these
cores, even if being far away, are connected to the central hub.
Similarly, the northern core seems to be connected to the main hub
trough filaments F01 and F02. Despite filament F01 being affected by
the \ion{H}{ii} region H01, the emission of \ce{CH3OH} reveals, through
a second velocity component, a link between the northern core and the
hub (see Fig. \ref{fig:PV-Plot}). Filament F02, which is only visible
in the emission of \ce{H2CS}, shows a smooth velocity gradient from
highly red to blue-shifted velocities when moving from the northern
core to the central hub (see Fig.~\ref{fig:H2CS_F02}). This suggests
that the distant cores as well as the northern core may fall onto the
hub, increasing the gas and stellar density, and favoring \SgrB(N) to
become a super stellar cluster.
\subsection{High-mass cluster formation}\label{sec:cluster_formation}
The high densities found in \SgrB(N) (10$^9$~cm$^{-3}$, or
10$^7$~$M_{\odot}$~pc$^{-1}$, see Paper II), the large mass reservoir of
the envelope, and high star formation
activity in a number of dense cores suggest that \SgrB(N) may gather
enough mass to form a super-stellar cluster or young massive cluster
(YMC), therefore constituting a good candidate YMC progenitor. The
formation of YMCs is still poorly understood, and different scenarios
have been proposed (e.g.\ \citealt{PortegiesZwart2010, Longmore2014}
for a review, and \citealt{Bressert2012, Fujii2015, Fujii2016,
Howard2018}). The "in-situ formation" scenario proposes that the
required amount of gas mass is gathered into the same volume of the
final stellar cluster before star formation sets in. The extreme
densities required to form a YMC, entail that the accumulated gas has a
short free-fall time, which in turn implies that the time for gathering
the final mass has to be short or, alternatively, the star formation
activity is delayed until the
accumulation of the total mass. In a second scenario, the "conveyor
belt formation" theory proposes that the mass reservoir extends
farther away than the final stellar cluster size. Thus, stars can form
in regions with much lower densities and be transported toward the
center of the cluster, increasing the final density of the YMC.
Our analysis of the properties of \SgrB(N) suggests a mix of the two
scenarios. %, with a small tendency to the `in-situ' cluster formation models
On one side, the dense cores located within the filaments
already harbor stars, according to our simulations in the ballpark of 2000
(see Appendix \ref{append:M_Lum_Lum}),
before merging into the main hub. This supports
the idea of the conveyor belt models ("dry" merger), in which star forming
clusters are transported throughout the filaments toward the center of
the region. On the other hand, the stellar mass ratio is 50\%, that is,\
the
dense cores are still surrounded by dense gas while reaching
the central region. Although our statistics are small, we see the trend
that dense cores located closer to the central hub can be more massive
than in the outer regions, (see Fig.~\ref{fig:M_core_dist}) suggesting that
they
have acquired more mass on their way to the main hub. In addition, the
comparison of timescales listed in Table~\ref{Table_FreeFall} suggests that
the dense cores can reach the center before they have exhausted all the gas
to form stars. Finally, the
measured high mass accretion rates ($0.08$--$0.16$~M$_\odot$~yr$^{-1}$)
point to a short time necessary to accumulate a large mass reservoir,
as proposed by the in-situ ("wet" merger) formation scenarios. Assuming constant
accretion rates, the central dense hub, with
($25$--$10)\times10^3$~$M_\odot$, has been assembled in only 60--300
kyr. However, the fraction of stars is large ($\sim50$\%) compared to the
amount of gas mass in the cores. As a result we cannot consider this process
as a clear wet merger, but as a "damp" merger. Overall, \SgrB(N) has the
potential to become a YMC or super stellar cluster similar to the Arches
and
Quintuplet clusters, which are also located in the CMZ near
the Galactic center.
%----------------------------------------------------------
\section{Summary}
We have studied the spatial distribution and kinematic properties of a
number of molecular species detected in the high-mass star-forming
region \SgrB(N). We used ALMA observations with an angular resolution
of 0$\farcs$4 (corresponding to 3300~au) and covering the frequency
range from 211 to 275~GHz. The rich chemistry of \SgrB(N) results in the
detection of thousands of lines,
often overlapping and therefore difficult to analyze. In
order to overcome this problem, we have developed a python-based tool
that stacks lines of any given molecular species resulting in an
increase of the final signal-to-noise ratio and averaging out line
blending effects. This has permitted us to study the distribution of
the molecular emission and its kinematic properties. Our main results
are summarized in the following:
\begin{itemize}
\item We have identified two different spatial structures related to
the molecular emission in \SgrB(N). Molecular species like
\ce{CH3OCH3}, \ce{CH3OCHO}, \ce{CH3OH}, or \ce{H2CS} show a filamentary
network similar to the distribution of the dust continuum emission at 242~GHz.
Contrary to that, species like
\ce{OCS} or \ce{C2H5CN} show a spherical or bubble-like distribution that
will be further analyzed in forthcoming papers.
\item We have identified eight filaments, which converge toward the central
massive core or hub. All but one filament are traced in multiple
molecules, including rare isotopologues. Filament F02 is detected in
our dataset in only one molecule, \ce{H2CS}. The filaments extent for
about 0.1~pc (some up to 0.5~pc) and have masses of a few
hundred $M_\odot$. The structure and distribution of the filaments,
together with the presence of a massive central region
($\sim$2000~$M_\odot$), suggest that these filaments play an important
role in the accretion process and transport mass from the outer regions
to the central hub.
\item From the line emission, we measure velocity gradients along the
filaments on the order of 20--100~km~s$^{-1}$~pc$^{-1}$ at scales of
$\sim$0.1~pc. This is 10--100 times larger than typical velocity
gradients found in other star forming regions at larger scales
($\sim$1~pc). We derive mass accretion rates in individual filaments of
up to 0.05~$M_\odot$~yr$^{-1}$, which add to a total of
0.16~$M_\odot$~yr$^{-1}$ when considering the accretion of all the
filaments into the central hub.
\item The dense cores identified in the 242~GHz continuum emission map are
found distributed along these filaments. We have determined the stellar mass
content of these cores, and compare the
timescale for the dense cores to collapse and form stars or clusters,
with the timescale required for them to be accreted onto the central
hub. Considering simple approximations, the free-fall timescale for
the core collapse is longer than the time of accretion onto the hub.
This suggests that although stars and clusters can form within the
cores while being accreted, they will not exhaust all the gas. This is
consistent with the stellar mass fractions of 50\% (with
respect to the total mass) that we derive. This suggests a scenario of a
damp merger \SgrB(N).
\item In summary, \SgrB(N) contains a central hub with large densities
($\sim$$10^9$~cm$^{-3}$, or $10^7$~$M_\odot$~pc$^{-3}$), a series of
massive dense cores ($\sim$200~$M_\odot$) with the potential to form
stellar clusters, and a network of filaments converging toward a central
hub,
with large velocity gradients and high mass accretion rates (up to
0.16~$M_\odot$~yr$^{-1}$). The large mass already contained in the
central hub ($\sim$2000~$M_\odot$ in about 0.05~pc size) together with
the merging process of dense cores already harboring stellar clusters,
suggest that \SgrB(N) has the potential to become a super stellar
cluster like the Arches or Quintuplet clusters.
\end{itemize}
% ---------------------------------------------
\begin{acknowledgements}
This work was supported by the Deutsche For\-schungs\-ge\-mein\-schaft (DFG)
through grant Collaborative Research Centre~956 (subproject A6 and C3, project
ID 184018867) and from BMBF/Verbundforschung through the projects ALMA-ARC
05A11PK3 and 05A14PK1. This paper makes use of the following ALMA data:
ADS/JAO.ALMA\#2013.1.00332.S.
ALMA is a partnership of ESO (representing its member states), NSF (USA)
and NINS (Japan), together with NRC (Canada) and NSC and ASIAA (Taiwan) and
KASI (Republic of Korea), in cooperation with the Republic of Chile. The
Joint ALMA Observatory is operated by ESO, AUI/NRAO and NAOJ.
DCL acknowledges support of the Scientific Council of the Paris Observatory.
Part of this research was carried out at the Jet Propulsion Laboratory, California
Institute of Technology, under a contract with the National Aeronautics and
Space Administration. In order to do the analysis and plots, we used the
python packages \texttt{scipy}, \texttt{numpy}, \texttt{pandas}, \texttt{matplotlib},
\texttt{astropy}, \texttt{aplpy} and \texttt{seaborn}.
\end{acknowledgements}
% --------------------------------------------------

%--------------------------------------------------------
\begin{appendix}
\section{Line stacking method}\label{append:line_stacking}
\begin{figure*}[ht!]
\centering
\includegraphics[width=0.9\textwidth]{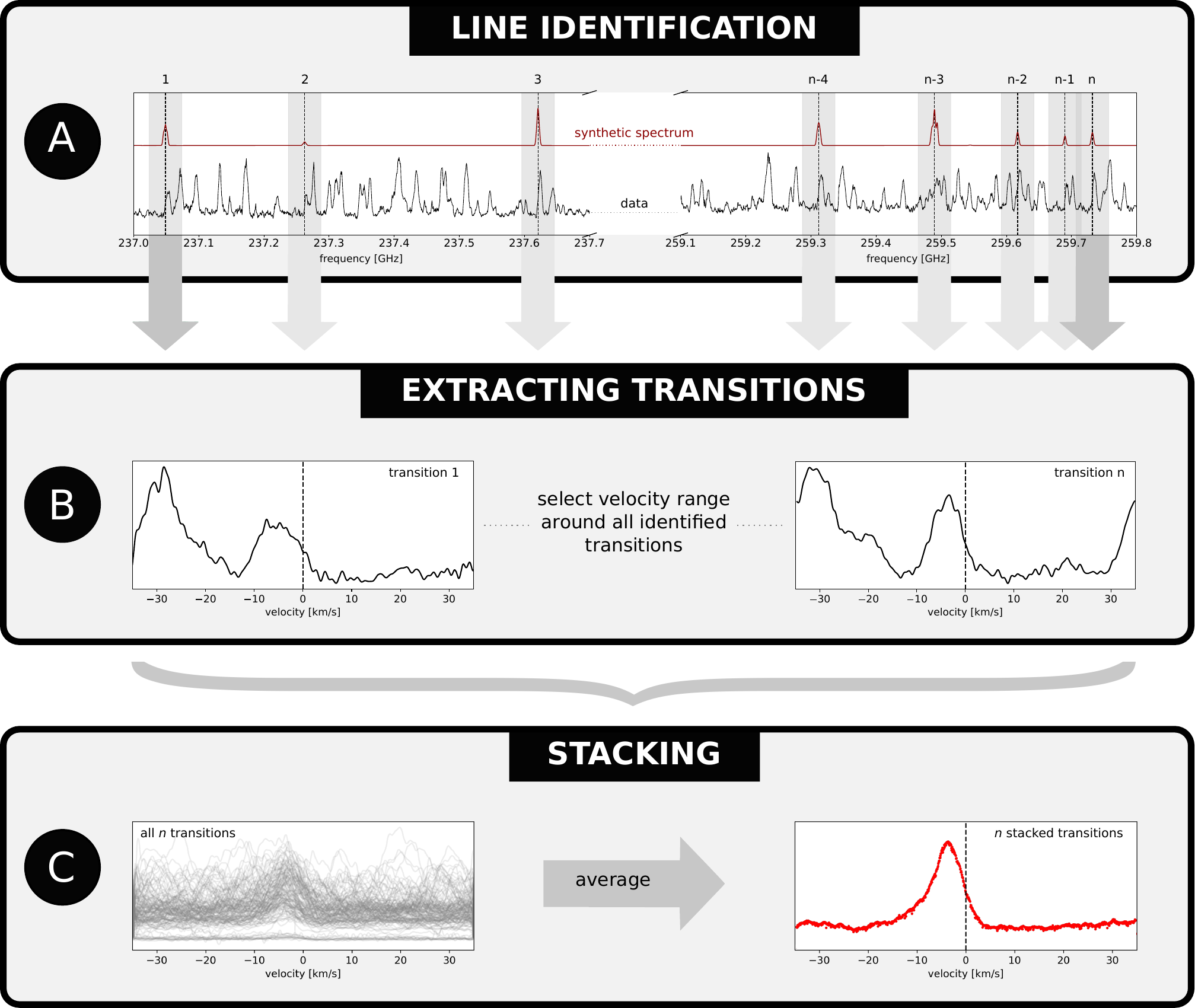}
\caption{Line stacking process for molecular lines in a
spectral survey. \textbf{Step~A}:
Correct observed spectrum,
shown in black in the top, by the source LSR velocity
(corresponding to 64~km~s$^{-1}$ for \SgrB(N)). Create a
synthetic spectrum (shown in red, and using XCLASS) for the
considered molecular species. This will be used for the
identification of detectable transitions (highlighted with
a
gray shadow area). \textbf{Step~B:} Extract a portion of
the
spectrum centered at each transition, and transform the
frequency axis into velocity using the Doppler equation.
Each
cutout spectra is uniformly resampled and rebined.
\textbf{Step~C:} Sum up all the cutout spectra and divide
by
the number of transitions. Normalization has been waived
to
prevent that noise and lines from other molecular species
are
not scaled up. Line blendings are statistically averaged
out
and the true line shape and the velocity can be determined.}
\label{fig:line_stacking}
\end{figure*}
The analysis of line-rich surveys with often thousands of lines has its
pitfalls due to an unknown number of spectral line components and
blending of multiple species. Particularly, deriving velocity
information is therefore rather complicated in line-crowded spectra. In
chemically rich regions like \SgrB, it is not easily possible to
identify well isolated transitions, which are usable to determine the
kinematic structure. Moreover, large velocity gradients across the
observed region as well as different chemical compositions throughout
the source are obstructive. In spite of these adversities, we convert
the difficulty of the presence of innumerable lines within a blind
survey into an advantage. For this, we have developed an automatized
(python based) tool to stack all transitions of a certain species to
increase the signal-to-noise ratio and to average out line blending
effects. The process of stacking is commonly used in Astronomy to
increase the signal-to-noise ratio of for example, weak radio galaxies,
recombination lines and even faint molecular line transitions (see
e.g., \citealt{Beuther2016, Lindroos2015, Loomis2018}). In our case,
stacked data also simplifies the structure and shape of the spectral
lines, thus making it possible to analyze the velocity and kinematic
properties.
The method is carried out in three main steps, illustrated in
Fig.~\ref{fig:line_stacking}. First (step~A in the figure), we correct
our observed data by the source LSR velocity, corresponding to about
64~km~s$^{-1}$ for \SgrB(N). Subsequently, we produce a synthetic
spectrum with XCLASS \citep{Moeller2017} for any molecular species
assuming similar conditions (i.e., temperatures, column densities) to those
representing the region (M\"oller et al.\ in prep). An
accurate fit of the observed data to derive the best temperature,
column densities and velocities is helpful but not mandatory.
Reasonable assumptions are enough to determine which frequency
transitions are detected in the observations, since the synthetic
spectrum is used for the identification of bright transitions. For
example, for the case of \ce{CH3OCHO}, we identified 250 bright lines
above a 10~$\sigma$ threshold in the frequency range 211 to 275~GHz.
This selection prevents us from including transitions that are included
in the catalog entries of the CDMS and JPL databases, but are too weak
and most likely non detected, in our source of interest. Moreover, the
exclusion of the weak transitions, reduces the addition of components
highly dominated by noise or by the presence of (bright) transitions
from other species.
Second (step~B in Fig.~\ref{fig:line_stacking}), we cut out all the
identified transitions and produce subspectra centered at them. These
spectra are transformed from frequency to velocity by applying the
Doppler equation. The velocity at 0~km/s (see
Fig.~\ref{fig:line_stacking}) corresponds to the rest frequency listed
in the database, since we have corrected the data by the source LSR
velocity. The spectra are uniformly resampled and rebined, using the
python packages \texttt{signal} and \texttt{numpy}, to ensure that the
channels in the velocity frame have the same width.
Finally (step~C in Fig.~\ref{fig:line_stacking}), all spectra are averaged
by taking the
the arithmetic mean. In the
final stacked spectra, the signal-to-noise ratio has increased, while
at the same time line blendings are averaged out. This simplifies
ensuing the analysis of the kinematic properties like velocity,
linewidth, lineshape or number of components.
The stacking process can
be applied to a subsample of transitions, selected for example, by excitation
energy level ($E_\mathrm{upper}$) or Einstein $A$ coefficient,
parameters included in the molecular line databases CDMS and JPL. This
selection of transitions will be used in a forthcoming paper to
characterize the excitation and physical conditions of the filaments in
\SgrB(N).
We performed tests with synthetic spectra to demonstrate the reliability
of the line stacking method. We have used the XCLASS software to generate
a synthetic spectrum of the molecular species C$_2$H$_3$CN (Acrylonitrile),
which is a good candidate since it has approximately 300 clearly detectable
transitions in the observed frequency range (211--275~GHz). The synthetic
spectrum of C$_2$H$_3$CN is produced including two different velocity components
with velocities of $0$ and $-5$~km~s$^{-1}$, and linewidths of $4$ and $2$~km~s$^{-1}$.
In Fig.~A.2, we show the results. In the top panel we have stacked the transitions
of the synthetic spectrum. The stacked spectrum shows a clear two-velocity
component profile. We fit Gaussian distributions and recover the velocities
and linewidths with an accuracy of 98\%. In a second run (see Fig.~A.2, bottom
panel) we combine the synthetic spectrum with the observed data of \SgrB(N).
We note that the molecule C$_2$H$_3$CN is not detected in our data and therefore
the only transitions that should be found are those include in the synthetic
spectrum. After performing line stacking, we recovered the velocities of
the two components with a high accuracy (about 98\%). The derived linewidths
are slightly larger than the original ones, resulting in an accuracy of 82\%.
This increase in the linewidth is likely produced by the presence of neighboring
transitions of other molecular species, which act like noise and result in
the broadening effect. Therefore, line-stacking of chemically rich regions
results in slightly broader lines. We also note that the presence of this
additional transitions artificial increases the continuum level, while not
affecting the velocity information of the lines.
\begin{figure}
\includegraphics[width=0.9\columnwidth]{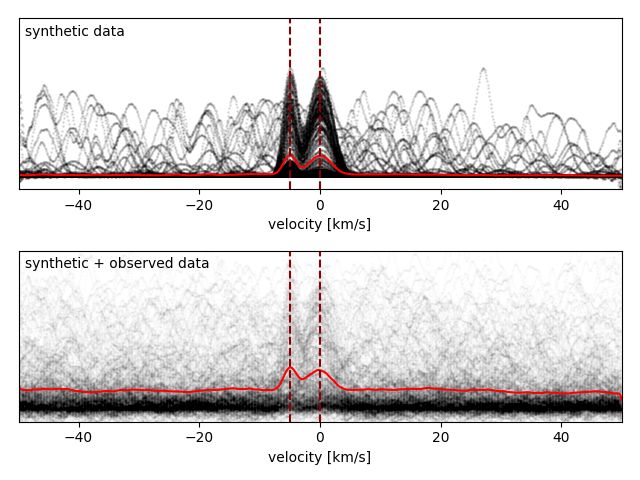}
\caption{Accuracy check of the line stacking method. \textit{Upper
panel}: Stacking performed with a pure synthetic spectrum of the molecular
species \ce{C2H3CN}. The velocity and velocity width have been fully recovered.
\textit{Lower panel}: Stacking of the same synthetic spectrum combined with
observed data. The line velocity stays recoverable with high precision, while
the velocity width has slightly increased but in a negligible order.}\label{fig:line_stacking_test}
\end{figure}
\section{Peak intensity maps}
\begin{figure*}
\includegraphics[width=0.9\textwidth]{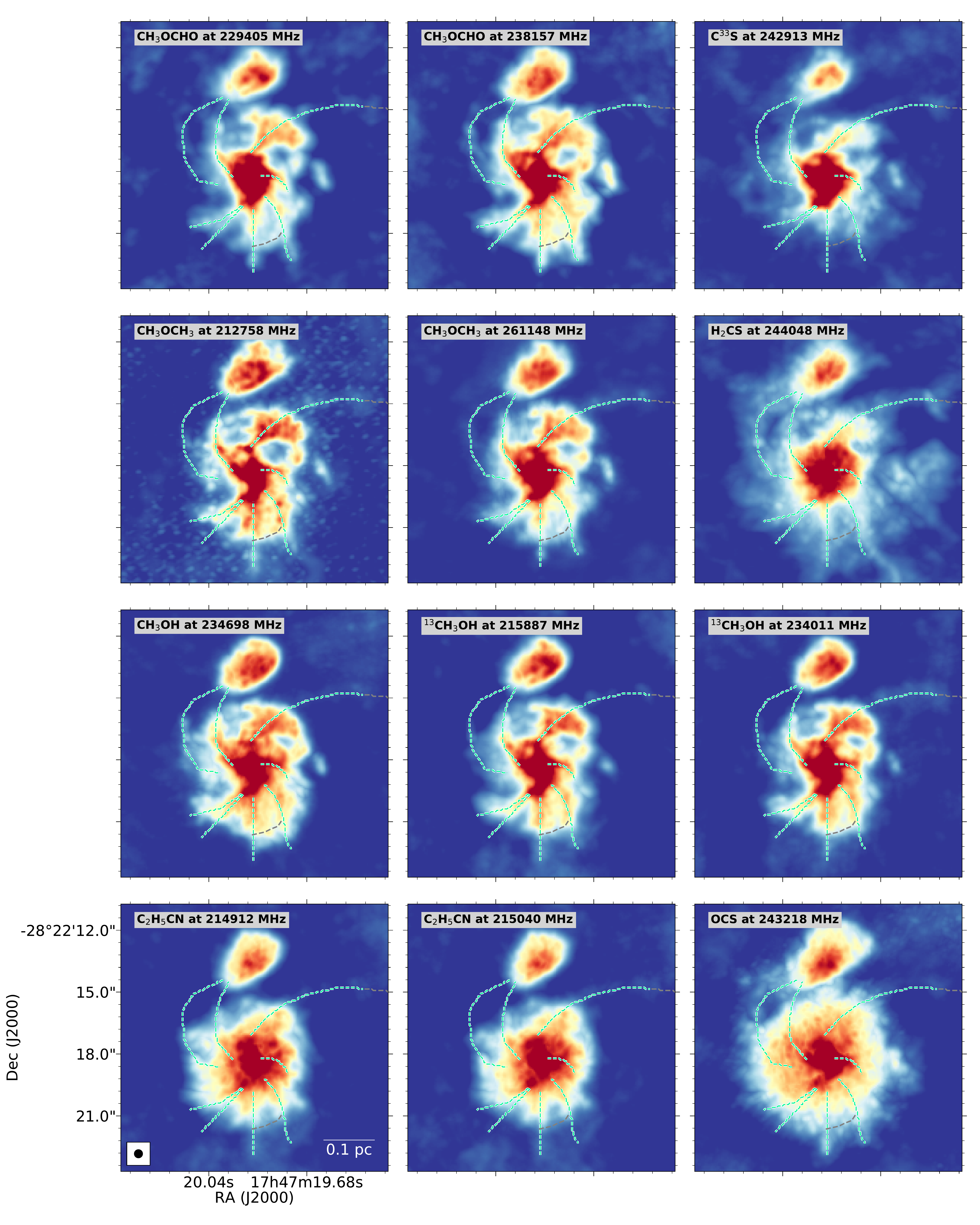}
\caption{Peak intensity maps of transitions of various molecular
species showing either the filamenary structure or bubble shape. The green
dashed lines trace the path of the filaments identified in the stacked data
(see Section~\ref{sec: Identification_properties}). }\label{fig:peak_int_apx}
\end{figure*}

In Fig.~\ref{fig:peak_int_apx} we present peak intensity maps of selected
transitions of different molecular species. The path of the filaments is
overlaid on all of them. The filamentary structures are clearly visible in
molecular species \ce{CH3OCHO}, \ce{CH3OCH3}, \ce{CH3OH,} and \ce{^13CH3OH}.
For other species such as \ce{C2H5CN} and \ce{OCS}, the bubble-shape structure
dominates the emission.
\section{Position-velocity plots along the filaments}\label{append:pv_plots}
\begin{figure*}
\centering
\includegraphics[width=0.9\textwidth]{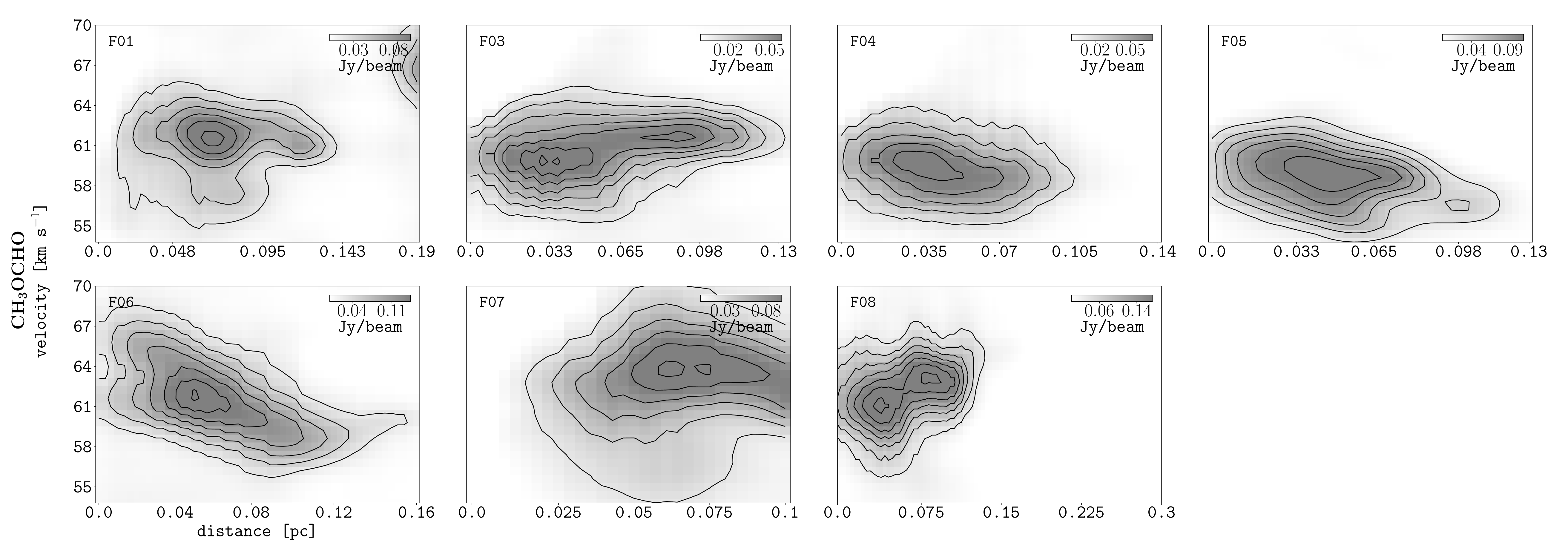}
\includegraphics[width=0.9\textwidth]{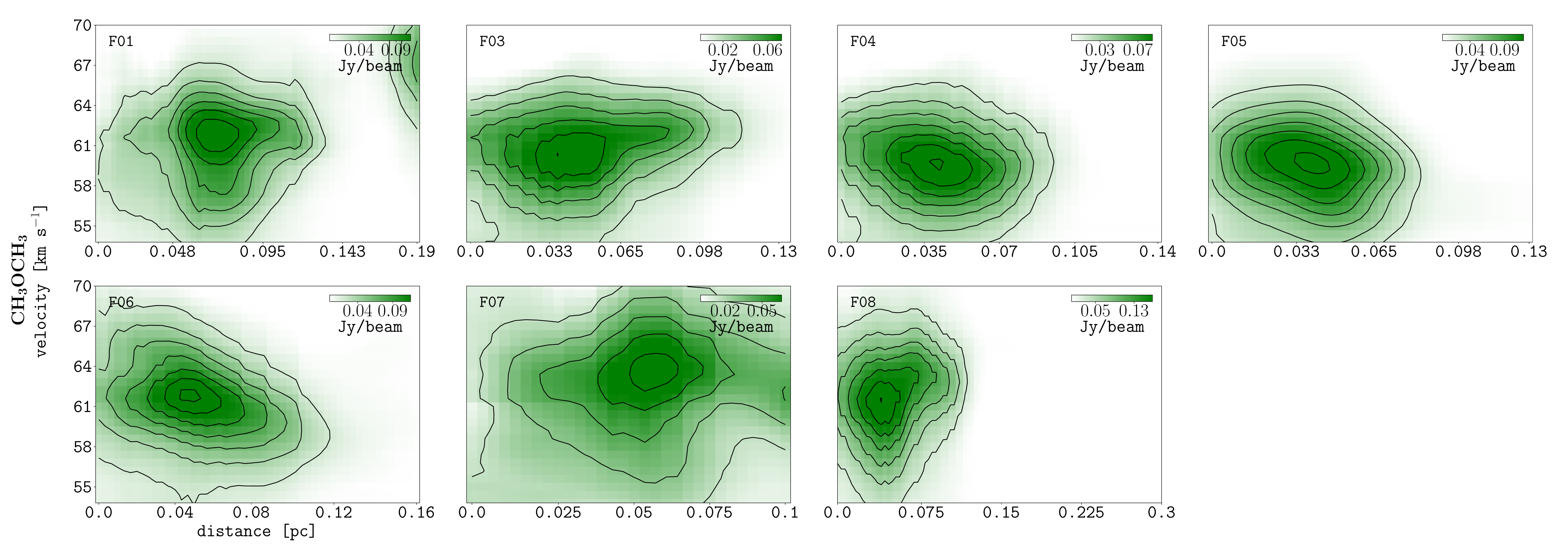}
\includegraphics[width=0.9\textwidth]{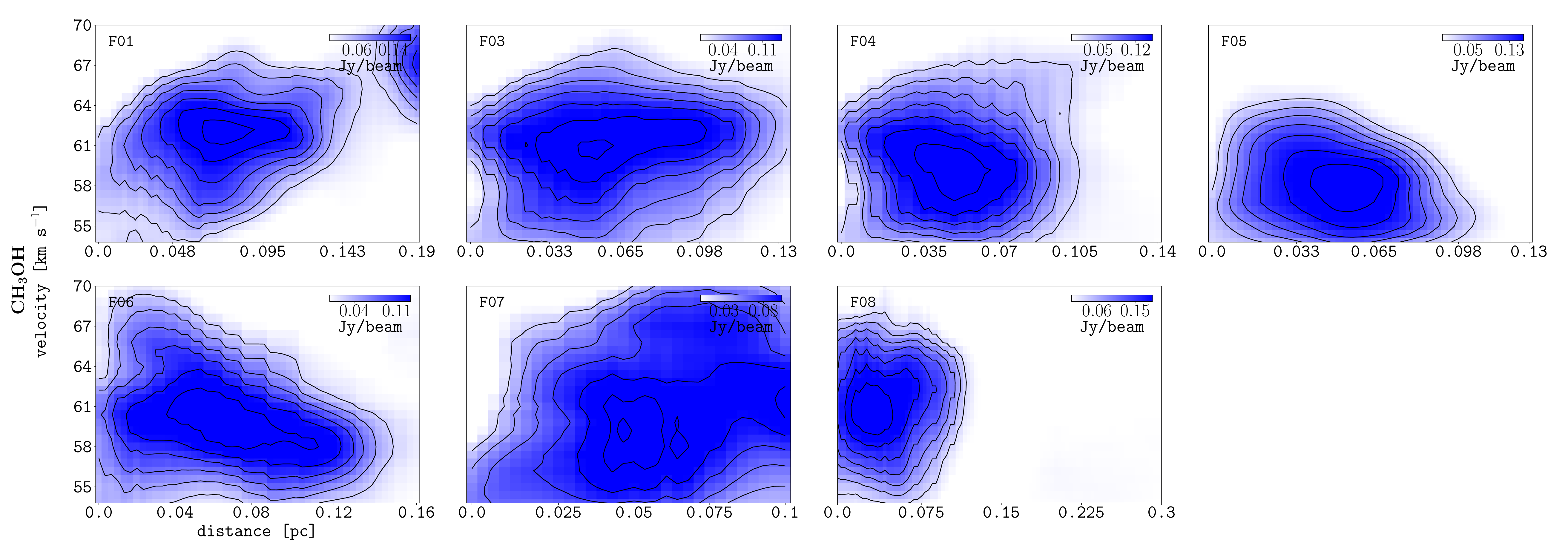}
\includegraphics[width=0.9\textwidth]{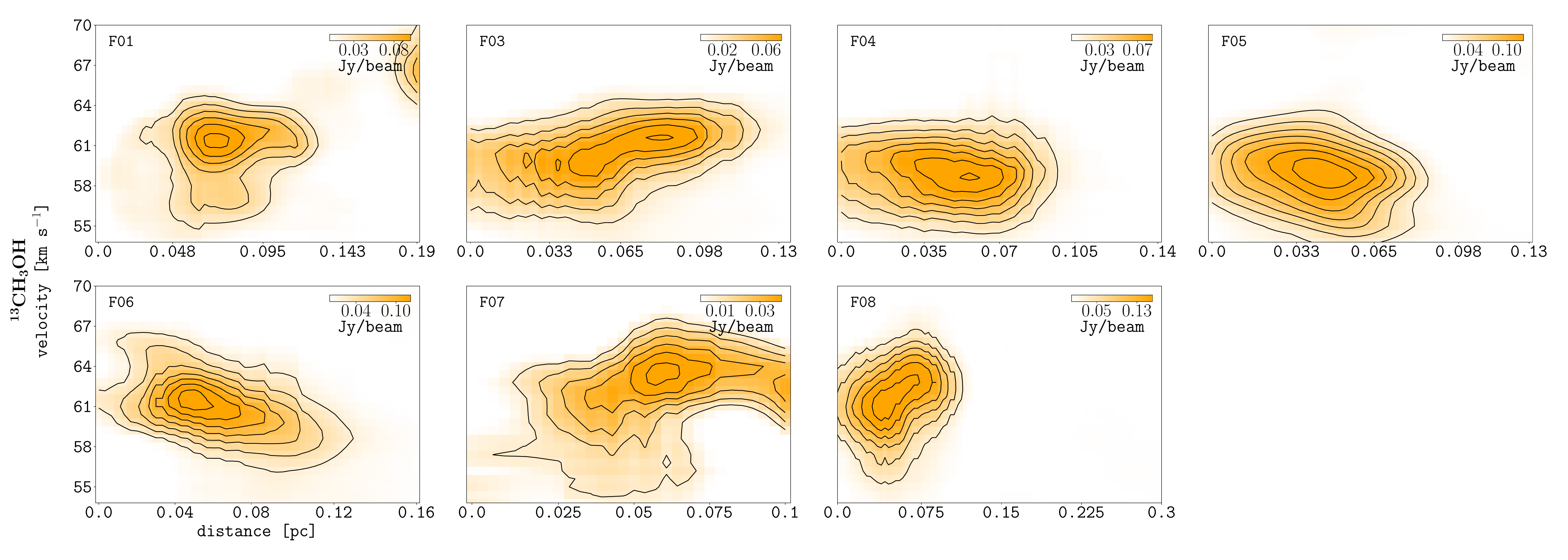}
\caption{Position-velocity plots along each filament, for the molecules
\ce{CH3OCHO} (gray), \ce{CH3OCH3}(green), \ce{CH3OH} (blue), and \ce{^13CH3OH}
(orange), from top to bottom.} \label{fig:PV-Plot}
\end{figure*}

\begin{figure}
\centering
\includegraphics[width=0.98\columnwidth]{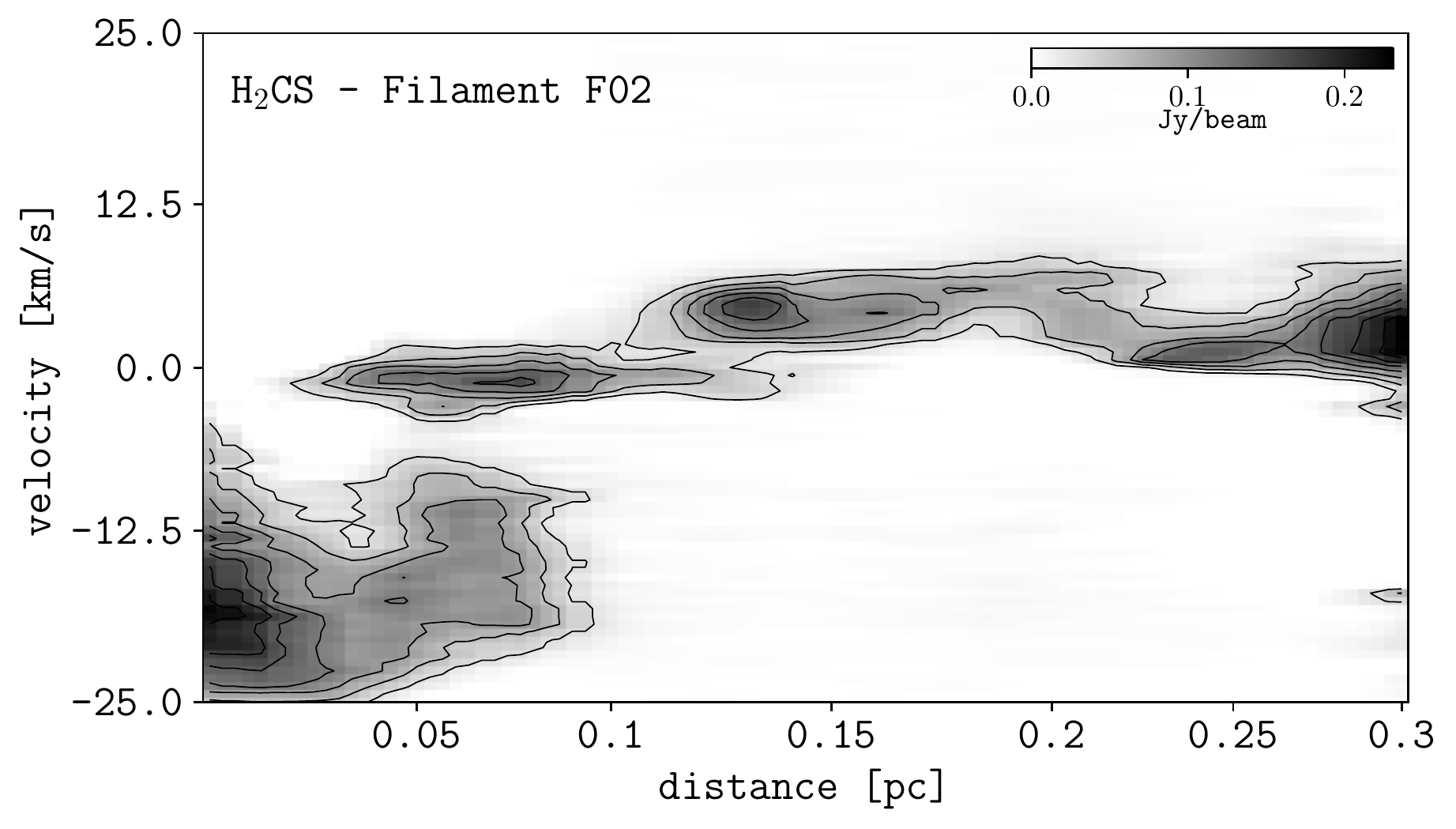}
\caption{Position velocity cut along filament F02 in \ce{H2CS}. The
emission to the left corresponds to the filament close to the central hub.
The emission at about 0.27~pc corresponds to the northern satellite core.}
\label{fig:H2CS_F02}
\end{figure}

Figure~\ref{fig:PV-Plot} shows the position-velocity plots produced
along the main spines of the filaments as depicted in
Fig.~\ref{fig:maps_stacked_cubes}. We have used the stacked cubes of
the molecular species \ce{CH3OCHO}, \ce{CH3OCH3}, \ce{CH3OH} and
\ce{^13CH3OH}, shown in gray, green, blue, and orange, respectively. The
position-velocity plots are obtained following a similar approach to
what is implemented in the python tool \texttt{pvextractor}, and
considering an averaging width across the filament of 0$\farcs$4,
corresponding to the synthesized beam. Similarly,
Fig.~\ref{fig:H2CS_F02} shows the position-velocity plot along filament
F02, which is only visible in \ce{H2CS}.
\section{Velocity gradients and linewidth variation}\label{append:velocity_gradients}
\begin{figure*}
\centering
\label{fig:violin_velocity}
\includegraphics[width=0.9\textwidth]{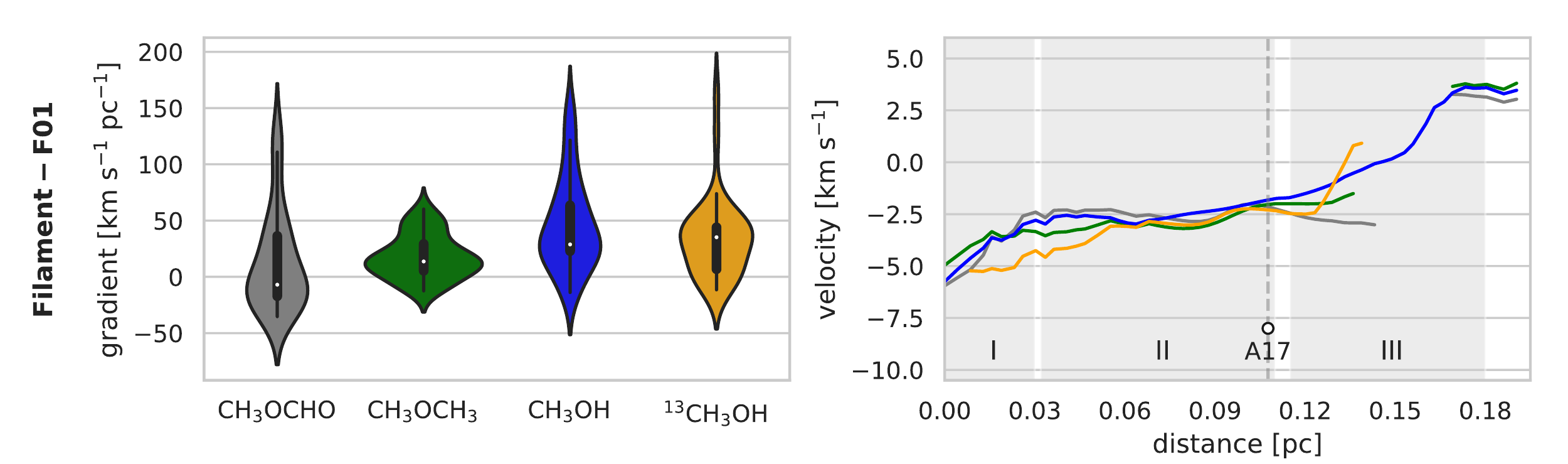}
\includegraphics[width=0.9\textwidth]{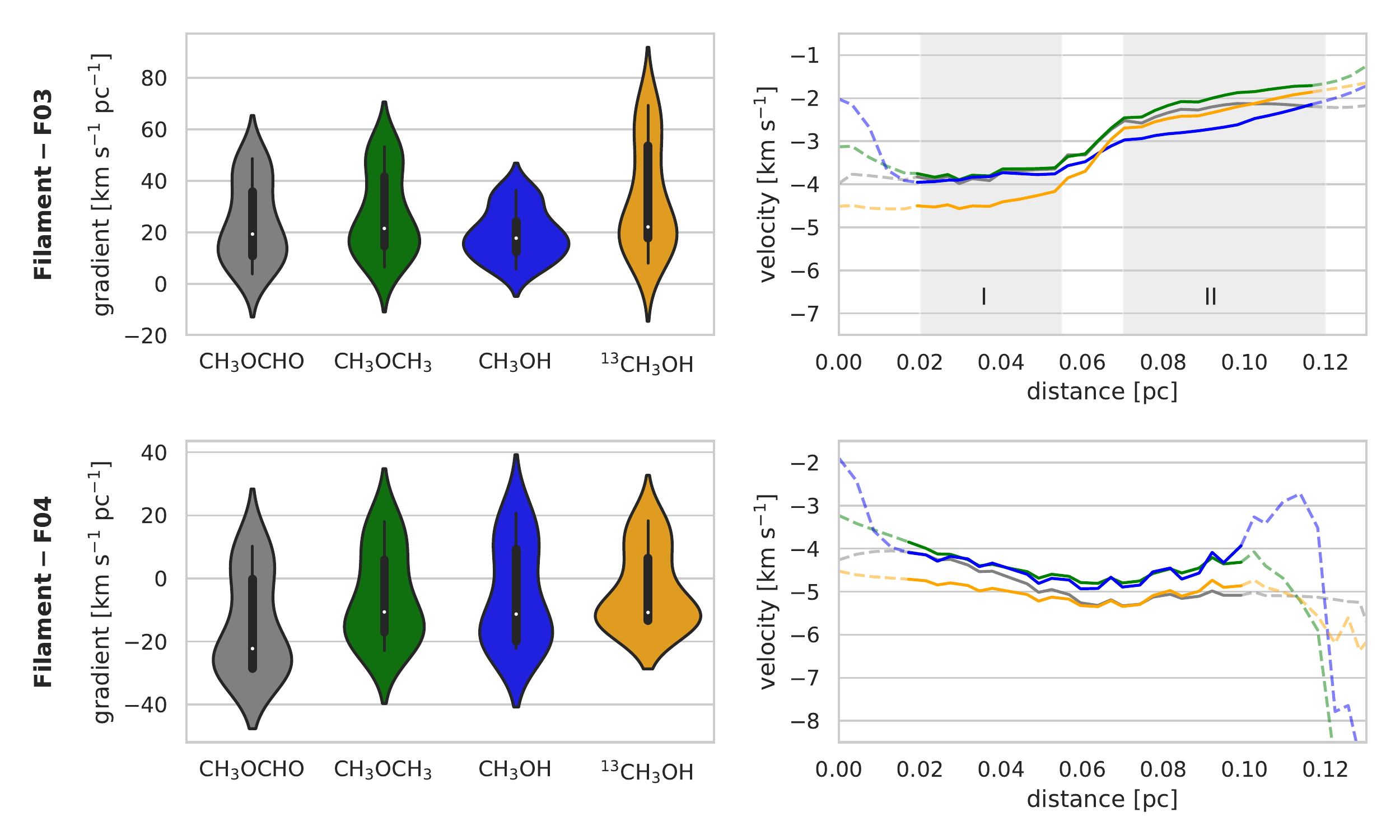}
\caption{\textit{Left panels:} Distribution of velocity gradients
along the filaments (determined in sections of 0.014~pc) shown in form of
a violin plot. The outer (or violin) shape represents all possible velocity
gradients, with broadening indicating how common they are, meaning that the
broadest part represents the mode average. The black thin bar indicates all
datapoints in the violin interior, while its thicker part corresponds to
the quartiles of the distribution, with the white dot indicating the mean
value of all velocity gradients along the respective filament. The distribution
is shown for the molecules \ce{CH3OCHO}, \ce{CH3OCH3} , \ce{CH3OH} and \ce{^13CH3OH}.
\textit{Right panel:} Velocity variation along the filaments. The gray areas
indicate subsections along the filaments, labeled with roman numbers. The
position of the dense cores are indicated by vertical, black dashed lines.
Regions where the line emission is below 4$\sigma$ are plotted with dashed
lines.} \label{fig:velocity_curve}
\end{figure*}

\begin{figure*}
\centering
\includegraphics[width=0.9\textwidth]{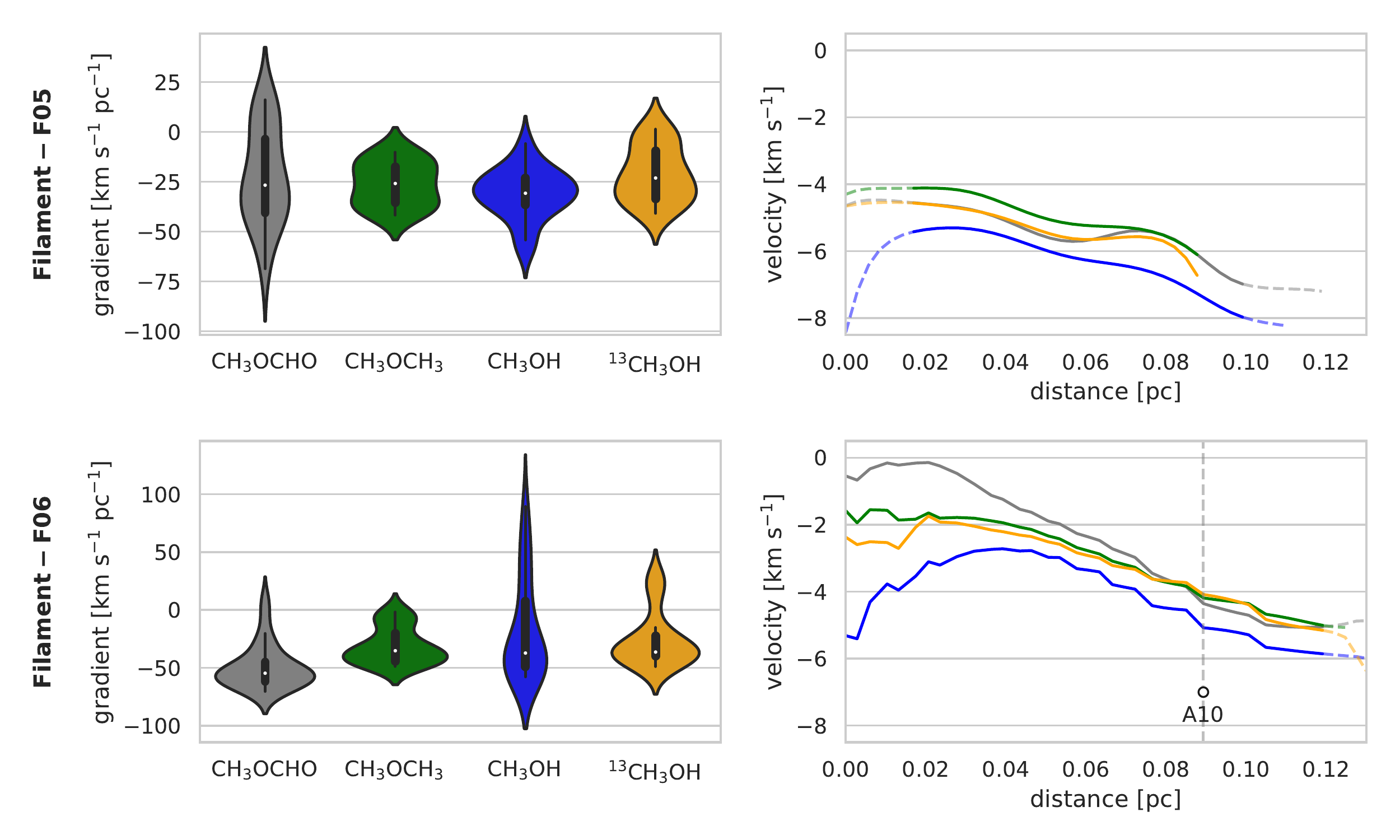}
\includegraphics[width=0.9\textwidth]{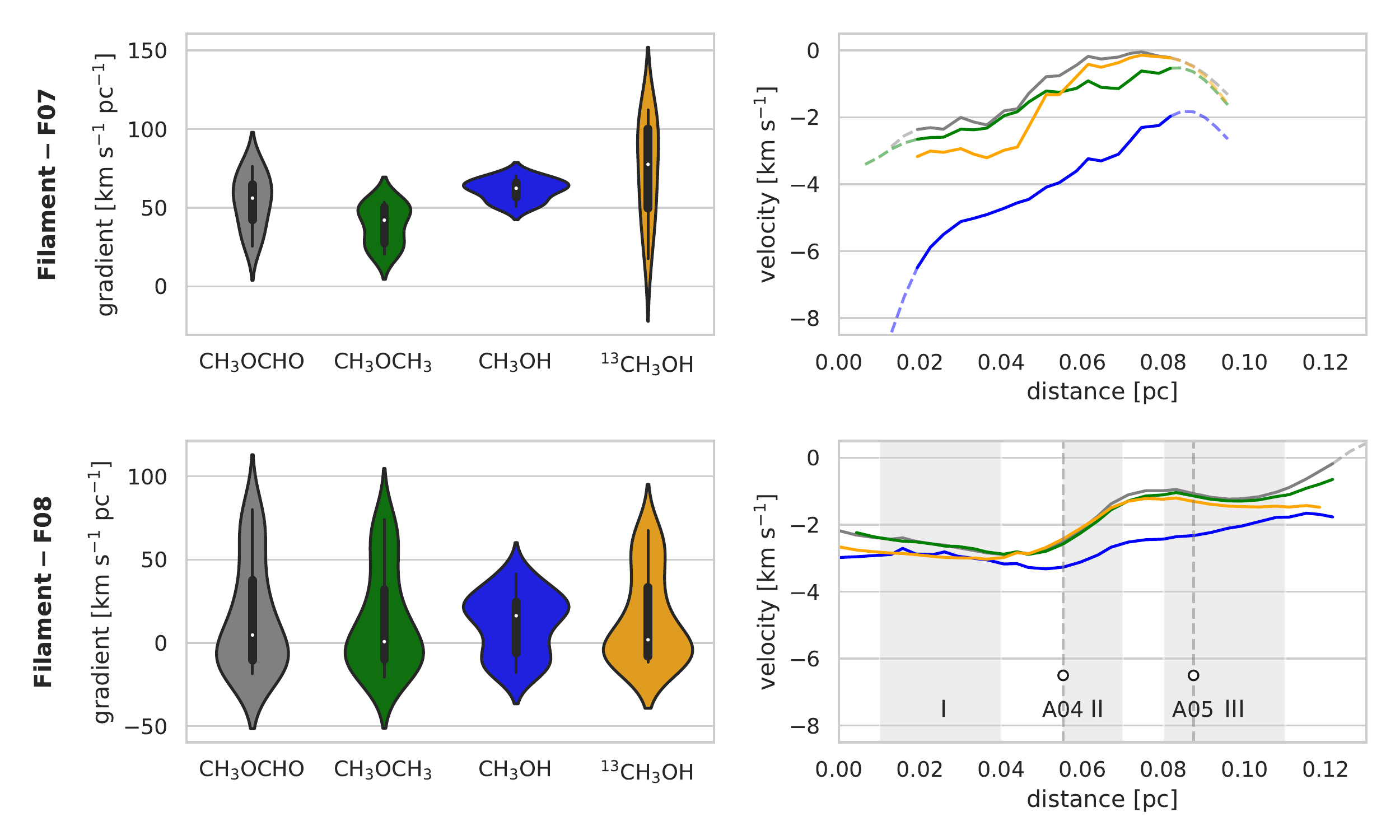}
\caption{Same as Fig.~\ref{fig:velocity_curve} for filaments F05,
F06, F07 and F08.} \label{fig:velocity_curve_2}
\end{figure*}

\begin{figure*}[ht!]
\centering
\includegraphics[width=0.9\textwidth]{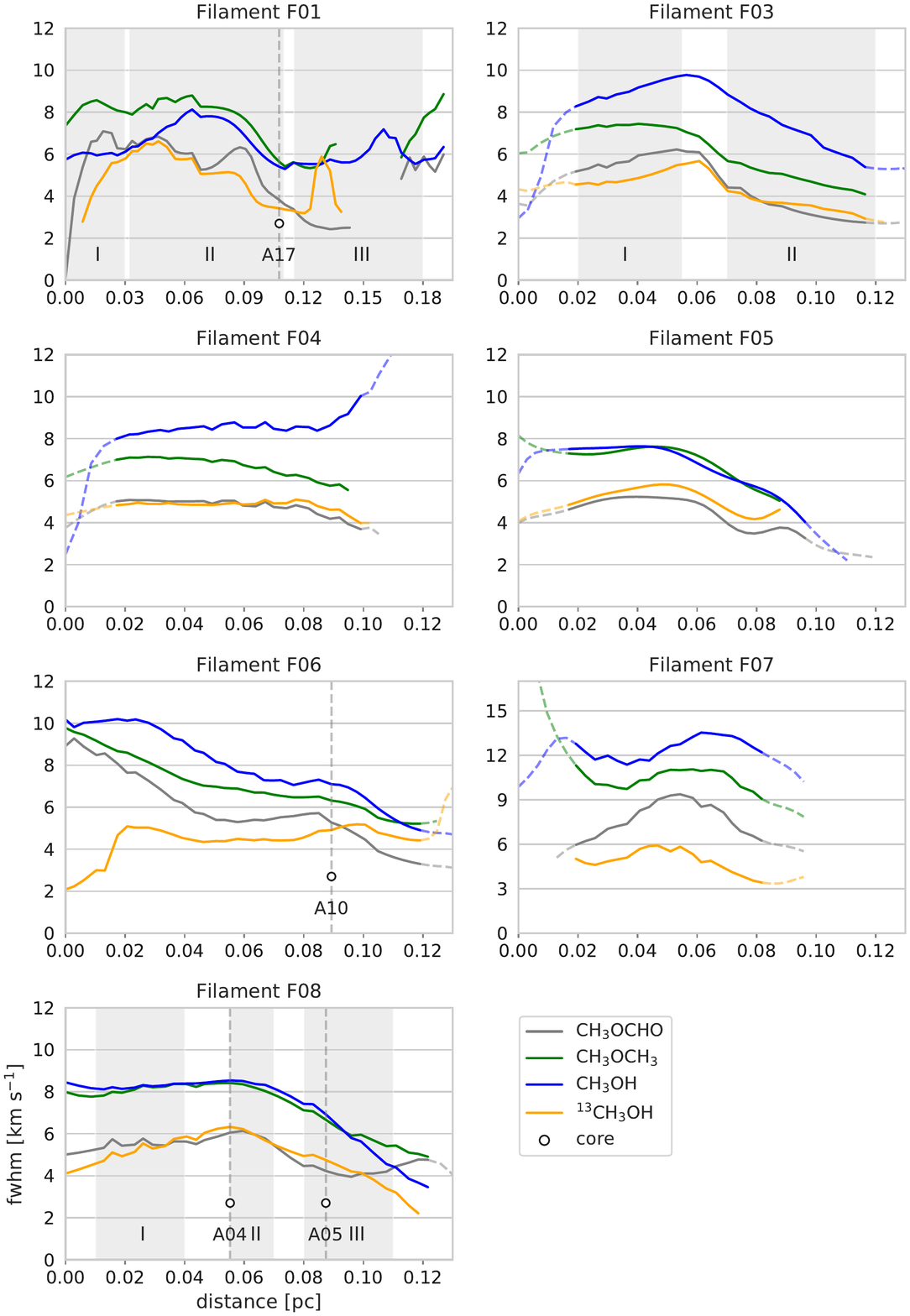}
\caption{Variation of the line width along the filaments. The position
of the dense cores are indicated by vertical, black dashed lines. Sections
for which the line emission is below 4$\sigma$ are plotted with dashed lines.}
\label{fig:fwhm_curve}
\end{figure*}

In order to constrain the velocity properties, we fit Gaussian functions
to the
spectra along the filaments to study the variation of velocity and
linewidth. In Figs.~\ref{fig:velocity_curve} and
\ref{fig:velocity_curve_2} (right panels), we plot the variation of
velocity with distance, while in Fig.~\ref{fig:fwhm_curve}, we plot the
variation of the linewidth (see Sect.~\ref{sec:results} for a
discussion on the kinematic properties of the filaments).
As discussed in Sect.~\ref{sec: filament_kinematics}, all
filaments show velocity gradients in the molecular species
\ce{CH3OCHO}, \ce{CH3OCH3}, \ce{CH3OH} and \ce{^13CH3OH}. We study the
variation of the velocity gradient along the filament by performing
linear fits in 0.014~pc size sections (corresponding to 0$\farcs$4, or
a beam size) along the filament. The distribution of the velocity
gradients are shown in the left panels of
Figs.~\ref{fig:velocity_curve} and \ref{fig:velocity_curve_2} in the
form of violin plots, which are a combination of box plots and a kernel
density plot. The distribution of velocity gradients for the different
filaments has, in most cases, a dominant peak, which is generally in
agreement with the average velocity gradient determined for the whole
filament (listed in Table~\ref{tab:filament_parameter}). The second
components in the distributions confirm the presence more than one
velocity gradient value along the filaments, suggesting a complex
velocity gradient structure which could be due to curvatures and
projection effects in the filaments.
\section{Blackbody assumption} \label{append:Black_body}
\begin{figure*}[!tbp]
\centering
\begin{minipage}[b]{0.48\textwidth}
\includegraphics[width=1\textwidth]{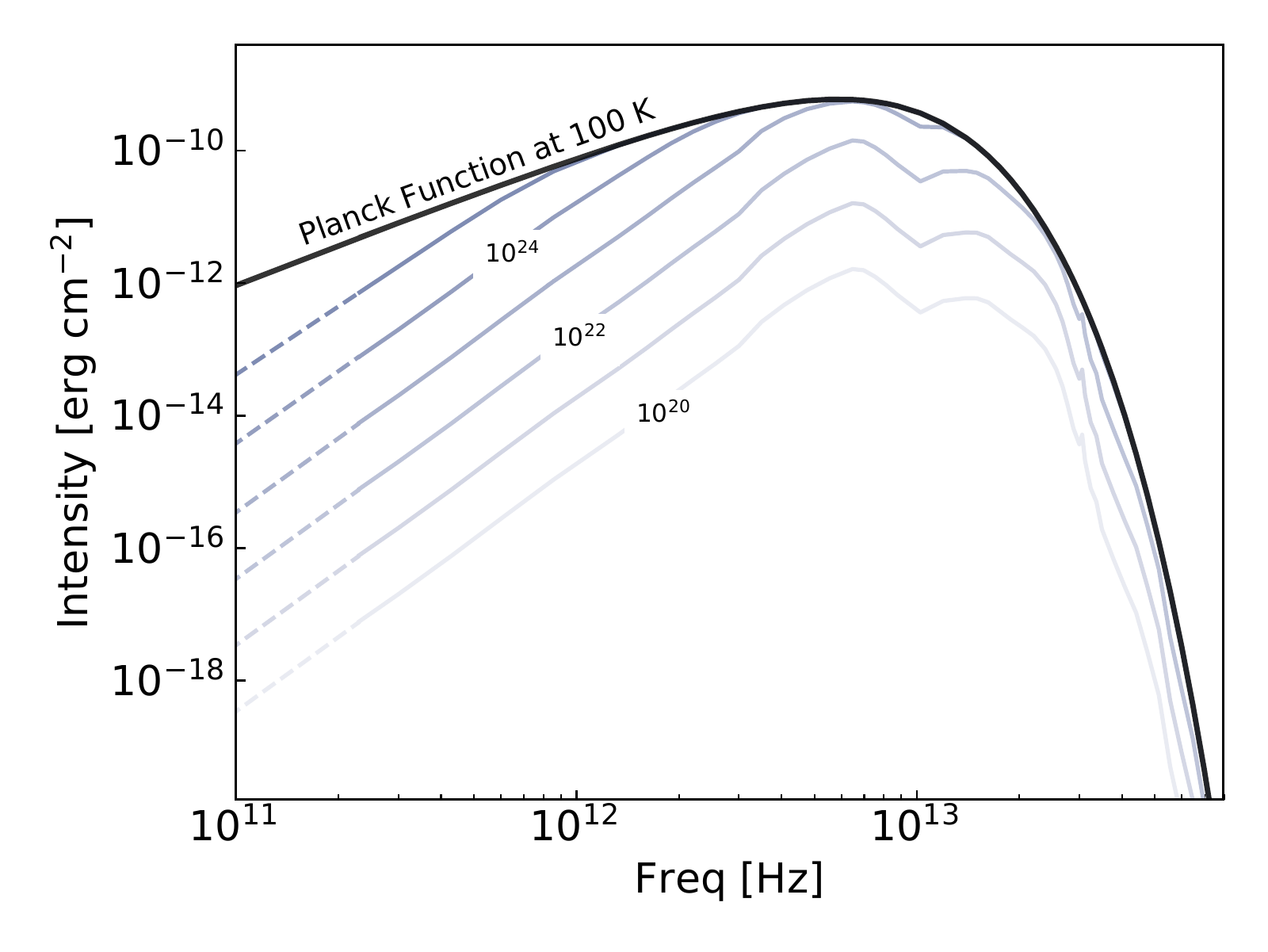}
\caption{Modified Planck function for different column densities
from 10$^{20}$ to 10$^{25}$~cm$^{-2}$
at a temperature of $100$~K. The solid gray curves
are calculated based on the opacities of \citet{Ossenkopf1994},
while the dashed lines are obtained from the extrapolation
of the two lowest frequency values in Ossenkopf \& Henning (1994). The Planck
function or perfect black body at a temperature of 100~K is shown with a
black solid and thick line.
} \label{fig:Planck_function}
\end{minipage}
\hfill
\begin{minipage}[b]{0.48\textwidth}
\includegraphics[width=1\textwidth]{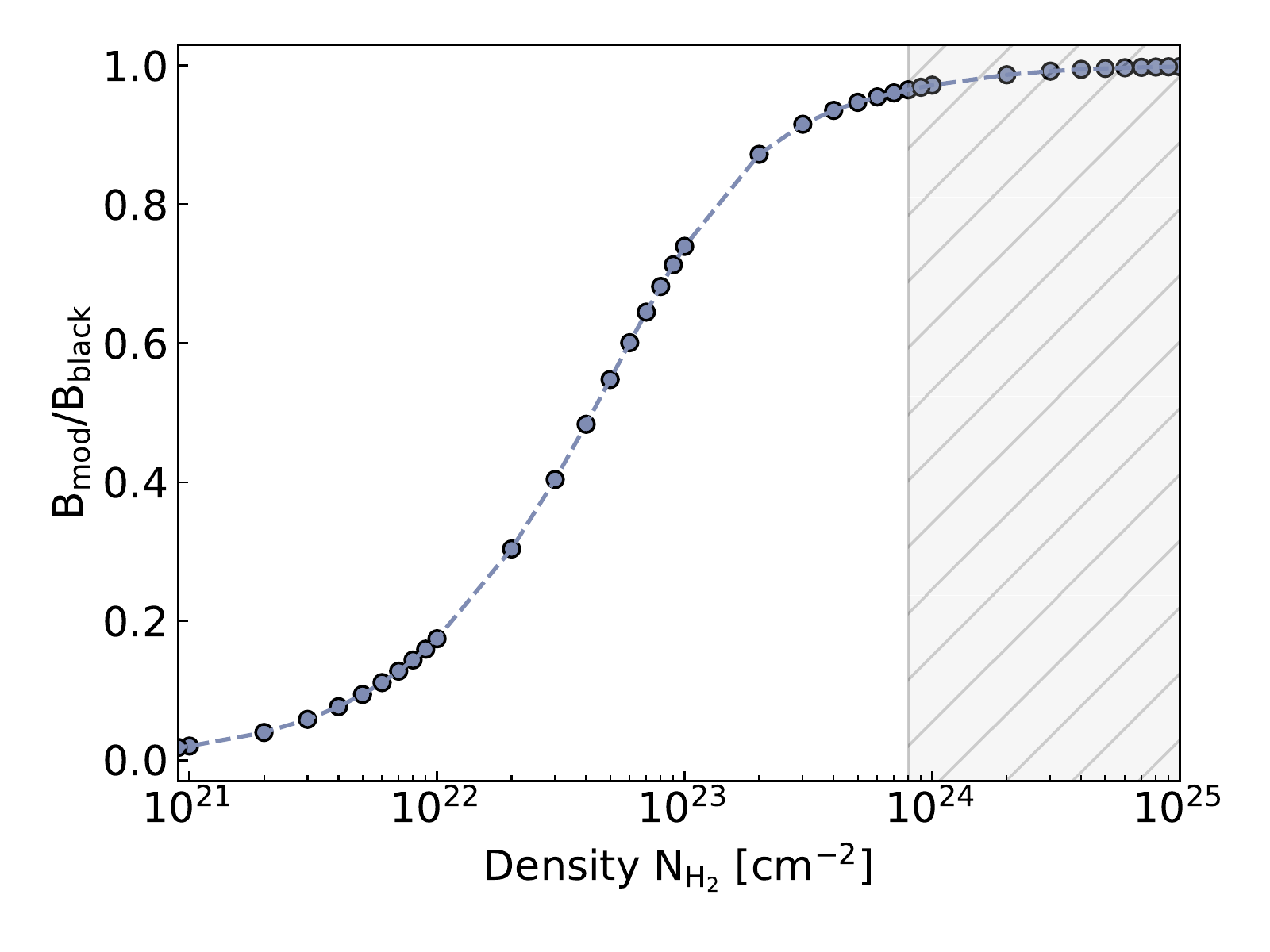}
\caption{Ratio between the total brightness of the modified
Planck function and a perfect blackbody. The gray dots marked the ratio for
specific column densities, while the dashed lines show the trend. The hatched
area indicates the range of column densities found in the dense cores of
\SgrB(N).} \label{fig:ratio-B-grey-black}
\end{minipage}
\end{figure*}

We derived the stellar luminosity of the dense cores assuming that they are
spherical black bodies and we use the Stefan-Boltzmann law. However, these
dense cores are not perfect black bodies, and the dust opacity has to be
taken into account. The modified blackbody or Planck function is
\begin{equation}
B_{\nu, \textrm{mod}}(T)= B_{\nu}(T) \, (1-e^{-\tau_{\nu}}),
\end{equation}
where $B_\nu(T)$ is the Planck function at temperature $T$ and $\tau_\nu$
is the optical depth at frequency $\nu$,
which is proportional to the density along the line of sight, given by
\begin{equation}
\tau_\nu = \int_\textrm{line of sight} \kappa_\nu \, \rho \, dl,
\end{equation}
with $\kappa_\nu$ being the absorption coefficient (opacity) per unit of
total mass density and $\rho$ the density.
We modeled the modified Planck function for different hydrogen column densities
$N_\mathrm{H_2}$ between
$10^{20}$--$10^{25}$~cm$^{-2}$ and a temperature of $100$~K (see
Fig.~\ref{fig:Planck_function}).
The absorption coefficients are taken from \citet{Ossenkopf1994} according
to
dust grains with thin ice mantles and gas densities of $10^6$~cm$^{-3}$.
We assume a dust-to-gas mass ratio of 100.
The total brightness of an emitting source can be obtained by integrating
over the (modified) Planck function
\begin{equation}
B(T) = \int_{0}^{\infty}B_\nu(T) \, d\nu. %\pi
\end{equation}
For a perfect black body, this results in
\begin{equation}
B(T)= \sigma T^4, \quad \sigma=5.67 \times 10^{-5} \textrm{erg cm}^{-2}
\textrm{s}^{-1}\textrm{K}^{-4},
\end{equation}
where $\sigma$ is the Stefan-Boltzmann constant. The luminosity calculated
from a numerical integration over the Planck function and multiplied by a
spherical surface of a given radius matches the luminosity calculated with
the Stefan-Boltzmann law. For a more detailed derivation see \citet{Wilson2009}.
In Fig.~\ref{fig:ratio-B-grey-black}, we present the ratio between the brightness
of a perfect black body and a modified black body as function of the column
density.
For hydrogen densities above roughly 5$\times$10$^{23}$~cm$^{-2}$, the total
brightness obtained for a modified Planck function is in agreement with the
brightness of a perfect black body, with deviations of less than 6\%. At
the observed high column densities, the emission is optically thick already
at lower frequencies compared to lower column densities (see deviations from
the Planck function in Fig.~\ref{fig:Planck_function}), and the variation
in the total integrated brightness is not significant.
In \SgrB(N) we have column densities about 10$^{24}$~cm$^{-2}$ \citep{Sanchez2017}.
Hence, the assumption of a blackbody is reasonable and the luminosity, which
is proportional
to the total brightness, can be calculated with the Stefan-Boltzmann law
(see Eq.~\ref{eq:stefan-boltzmann-law}).
\section{Stellar luminosity and mass relation for clusters}\label{append:M_Lum_Lum}
\begin{figure*}[!tbp]
\centering
\begin{minipage}[b]{0.48\textwidth}
\includegraphics[width=1\textwidth]{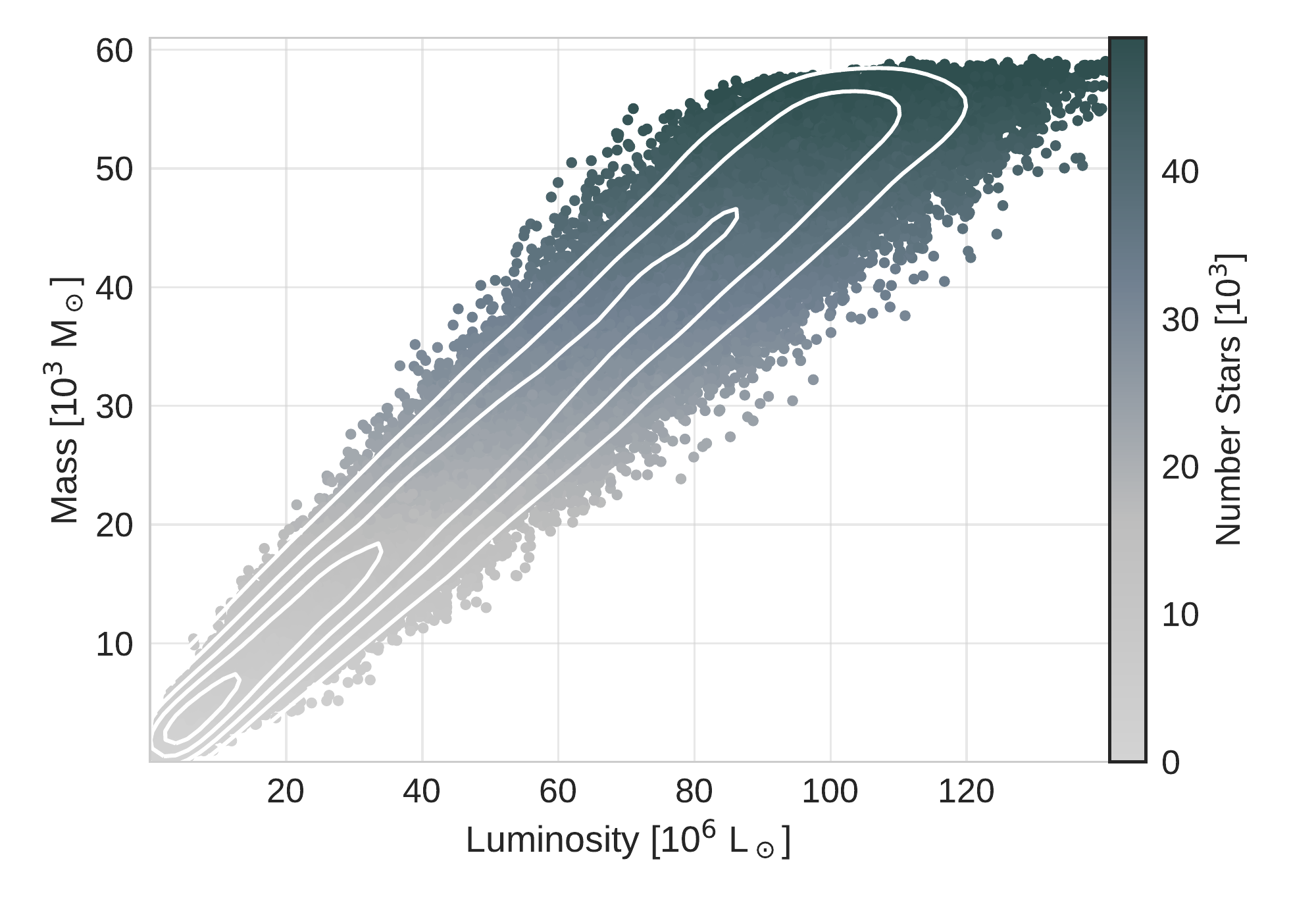}
\caption{Relation between the stellar mass and stellar luminosity
for clusters with different numbers of stars. The white contours show the
2D kernel density estimate (KDE) of the simulated clusters.
} \label{fig:Mass_Lum}
\end{minipage}
\hfill
\begin{minipage}[b]{0.48\textwidth}
\includegraphics[width=1\textwidth]{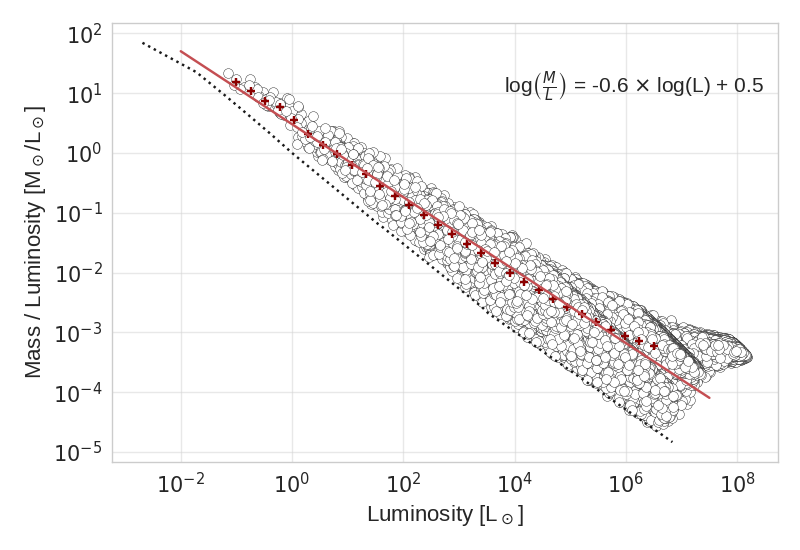}
\caption{Relation between the ratio $(M/L)_\mathrm{cluster}$
with the cluster luminosity. The white circles show all simulated clusters
and the red crosses are the mean values of $(M/L)_\mathrm{cluster}$ for different
luminosities. The red line indicates a linear fit through the mean values.
The black dashed line shows the relation between luminosity and mass for
a single star (see Eq.~\ref{eq:relation_lum_mass_star})} \label{M_Lum_Lum}
\end{minipage}
\end{figure*}

\begin{figure*}[ht!]
\centering
\includegraphics[width=0.95\textwidth]{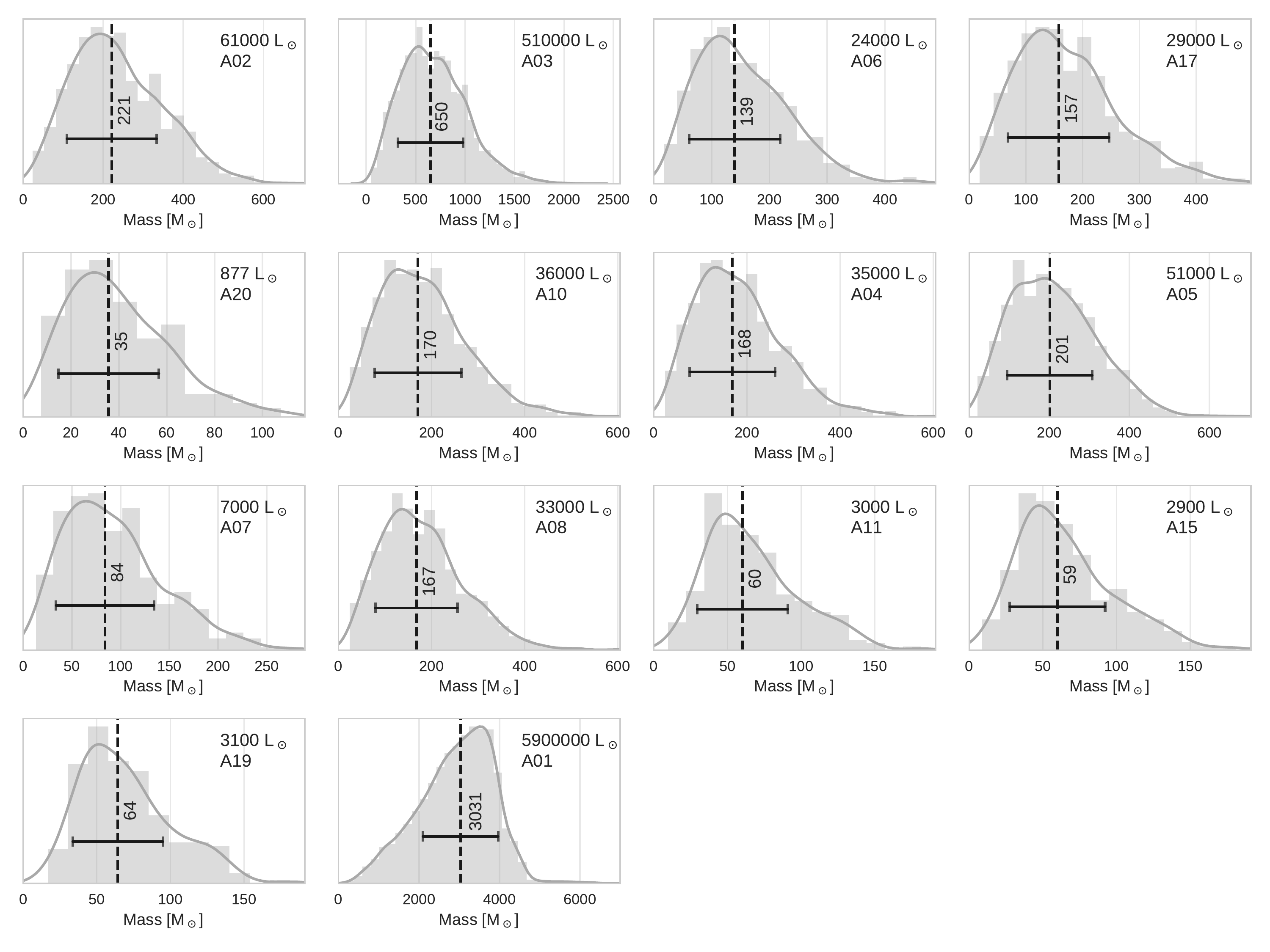}
\caption{Stellar mass distribution of clusters with luminosities
according to the observed dense cores in \SgrB(N). The gray line shows the
KDE obtained with the python package \texttt{seaborn}. The black dashed vertical
lines indicate the mean values of the mass distribution, and the horizontal
lines indicate the standard deviation.} \label{kde}
\end{figure*}

In order to determine the stellar mass and luminosity of clusters, we simulated
10$^5$ clusters with a random amount of stars. The number of stars varies
between 1 and $50000$, with masses from $0.1$ and $100$~M$_\odot$. The mass
distribution of stars for each cluster follows the Initial Mass Function
(IMF, \citealt{Kroupa2001}),
\begin{equation}
\xi(M) \propto
\begin{cases}
M^{1.3} & \text{if } 0.1~M_\odot < M \leq 0.5~M_\odot \\
M^{2.3} & \text{if } M > 0.5~M_\odot
\end{cases}
\end{equation}
where $\xi$ is the number of stars of a certain mass. The stellar luminosity
($L$) for a given star of mass $M$ follows the relation (see \citealt{Eker2018})
\begin{equation}\label{eq:relation_lum_mass_star}
\mathrm{log} \left( \frac{L}{L_\odot} \right) = a \cdot \mathrm{log} \left(
\frac{M}{M_\odot} \right) - b ,
\end{equation}
where the constants $a$ and $b$ vary depending on the stellar mass. For the
range of masses considered, we use (based on \citealt{Eker2018}):
\begin{align*}
0.1 < \mathrm{M}/M_\odot \leq 0.45 \qquad \mathrm{log }\, L = 2.028 \times
\mathrm{log }\, \mathrm{M} - 0.976 \\
0.45 < \mathrm{M}/M_\odot \leq 0.72 \qquad \mathrm{log }\, L = 4.572 \times
\mathrm{log }\, \mathrm{M} - 0.102 \\
0.72 < \mathrm{M}/M_\odot \leq 1.05 \qquad \mathrm{log }\, L = 5.743 \times
\mathrm{log }\, \mathrm{M} - 0.007 \\
1.05 < \mathrm{M}/M_\odot \leq 2.40 \qquad \mathrm{log }\, L = 4.329 \times
\mathrm{log }\, \mathrm{M} + 0.010 \\
2.40 < \mathrm{M}/M_\odot \leq 7.0\phn \qquad \mathrm{log }\, L = 3.967
\times \mathrm{log }\, \mathrm{M} + 0.093 \\
7.0\phn < \mathrm{M}/M_\odot \leq 100 \qquad \mathrm{log }\, L = 2.865
\times \mathrm{log }\, \mathrm{M} + 1.105.
\end{align*}
The masses and luminosities of all individual stars are summed, and result
in a total mass and luminosity per cluster. The result for all simulated
clusters is shown in Fig. \ref{fig:Mass_Lum}, presenting the relation between
the stellar mass of the cluster and the cluster luminosity. In Fig.~\ref{M_Lum_Lum},
we show the relation between $(M/L)_\mathrm{cluster}$ and the cluster luminosity.
We fit a linear function to the relation and find
\begin{equation}
\mathrm{log} \: \left( \frac{M/M_\odot}{L/L_\odot} \right) = -0.6 \times
\mathrm{log} \left( \frac{L}{L_\odot} \right) + 0.5,
\end{equation}
where $M$ is the mass of the cluster and $L$ the cluster luminosity.
In the particular case of the cores of \SgrB(N), we determine the stellar
mass of the clusters that form inside by making use of the luminosities listed
in Table~\ref{Table_FreeFall}. For this, we extract all the simulated clusters
with luminosities corresponding to the core luminosity, within a range of
$\pm$10\%. The stellar mass distributions of the selected clusters are shown
as histograms in Fig.~\ref{kde} We finally determined the stellar mass $M_\mathrm{stellar}$
as the median value of the distribution. We note that the number of stars
within the dense cores of \SgrB(N) is found to be below 2000 objects (see
also Fig.~\ref{fig:Mass_Lum})
\section{Mass-to-length ratio}\label{append:mass_to_length}
\begin{figure*}
\centering
\includegraphics[width=2\columnwidth]{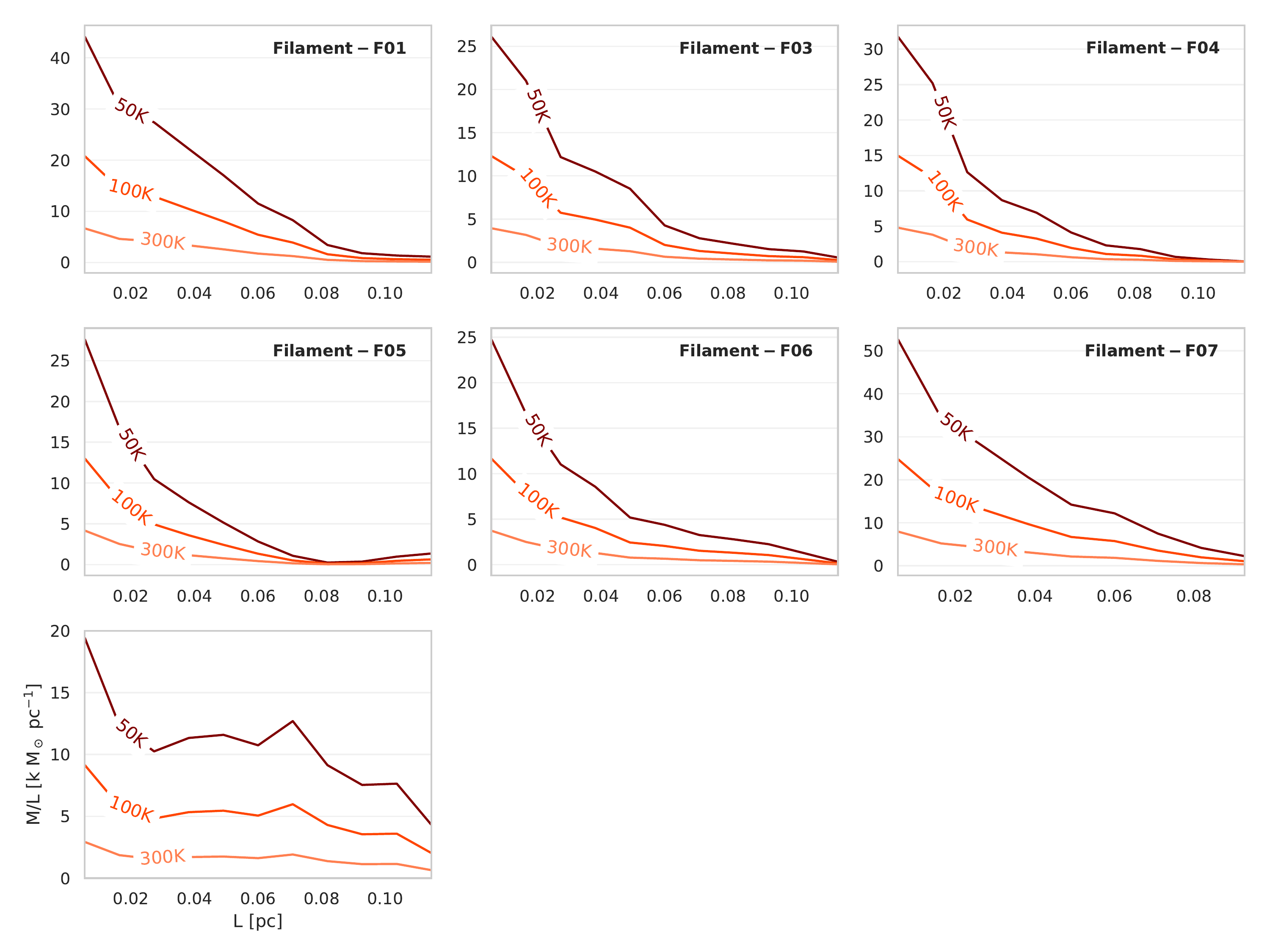}
\caption{Variation of mass-to-length ratio ($M$/$L$) along the filaments,
computed along sections of 0.011~pc in length. The different lines correspond
to different assumed temperatures: 50~K (dark), 100~K (orange), and 300~K
(light orange).} \label{fig:M/L_variation}
\end{figure*}

In Fig.~\ref{fig:M/L_variation} we plot the variation of mass-to-length
ration along the filament. The mass has been computed for segments of
0.011~pc in length and for three different dust temperatures: 50~K,
100~K, and 300~K. Taking the 50~K case as an example, we find for all
filaments an increase of mass when approaching the center of the
region. This may be due to a higher mass content in the inner regions
of the filaments. Alternatively, this might be an artifact produced by
a wrong temperature assumption. If the filament is hotter in the inner
regions (e.g., six times hotter, from 50~K to 300~K) the increase in mass
toward the center is less significant, although still existing in many
filaments.
\end{appendix}
\end{document}